\documentclass[article,longauth]{aa}

\usepackage[varg]{txfonts}
\usepackage{graphicx}
\usepackage[breaklinks=true]{hyperref}
\usepackage{placeins}
\usepackage{xcolor}
\providecommand{\srcfirst}{EP\,J223759.5$+$531421}
\providecommand{\src}{EP\,J2237+5314}
\providecommand{\nustar}{\textit{NuSTAR}}
\providecommand{\xmm}{\textit{XMM-Newton}}

\def\xmm{\emph{XMM--Newton}}

\def\nustar{\emph{NuSTAR}}

\def\swift{\emph{Swift}}
\def\ep{\emph{Einstein Probe}}

\def\flux{\mbox{erg\,cm$^{-2}$\,s$^{-1}$}}
\def\lum{\mbox{erg\,s$^{-1}$}}
\def\nh{\ensuremath{N_{\rm H}}}

\def\arcsec{\mbox{$^{\prime\prime}$}}
\def\arcmin{\mbox{$^{\prime}$}}
\def\deg{\mbox{$^{\circ}$}}

\begin{document}

\title{Einstein Probe discovery of the magnetar EP\,J223759.5$+$531421 }

\author{
Nanda~Rea\inst{1,2}
\corrauth{rea@ice.csic.es, cotizelati@ice.csic.es, Alice.Borghese@esa.int}
\and
Francesco~Coti~Zelati\inst{1,2,3,\star}
\and
Alice~Borghese\inst{4,\star}\thanks{ESA Research Fellow}
\and
Yilong~Wang\inst{1,2}
\and
Haonan~Yang\inst{5}
\and
Xuan~Mao\inst{5,6}
\and
Cui-Yuan~Dai\inst{7,8}
\and
E.~Arrigoni\inst{15}
\and
J.~Bai\inst{34}
\and
G.~Bernardi\inst{14}
\and
M.~Burgay\inst{13}
\and
S.~Cao\inst{5}
\and
A.~Coleiro\inst{24}
\and
D.~De~Grandis\inst{1,2}
\and
P.~Esposito\inst{12,15}
\and
H.~Feng\inst{25}
\and
Y.-C.~Fu\inst{17,18}
\and
A.~Geminardi\inst{12,13,35}
\and
D.~G\"otz\inst{16}
\and
S.~Guillot\inst{19}
\and
Y.~Huang\inst{33}
\and
M.~Imbrogno\inst{1,2}
\and
G.~L.~Israel\inst{11}
\and
C.~Jin\inst{5,6}
\and
A.~Kong\inst{20}
\and
Y.~Li\inst{31}
\and
L.~Lin\inst{17,18}
\and
C.-S.~Liu\inst{5}
\and
C.~Maitra\inst{21,23}
\and
A.~Marino\inst{1,2,26,27}
\and
S.~Mereghetti\inst{15}
\and
J.-U.~Ness\inst{4}
\and
S.~Ng\inst{22}
\and
M.~Pilia\inst{13}
\and
A.~Possenti\inst{13}
\and
H.~Sun\inst{5,6}
\and
L.~Tao\inst{25}
\and
R.~Taverna\inst{9}
\and
A.~Tiengo\inst{12,15}
\and
R.~Turolla\inst{9,10}
\and
W.~Yuan\inst{5}
\and
S.~Zane\inst{10}
\and
B.~Zhang\inst{22,28,29,30}
\and
L.-X.~Zhang\inst{32}
\and
Z.~Zhang\inst{17,18}
\and
W.-W.~Zhu\inst{5,17}
\and
Y.-C.~Zou\inst{31,32}
}

\authorrunning{Rea, Coti Zelati, Borghese et al.}

\institute{
Institute of Space Sciences (ICE--CSIC), Campus UAB, Carrer de Can Magrans s/n, 08193 Bellaterra, Spain
\and
Institut d'Estudis Espacials de Catalunya (IEEC), Edifici RDIT, Campus UPC,
E-08860 Castelldefels, Barcelona, Spain
\and
INAF--Osservatorio Astronomico di Brera, Via E. Bianchi 46,
I-23807 Merate (LC), Italy
\and
European Space Astronomy Centre (ESAC), ESA,
Camino Bajo del Castillo s/n, E-28692 Villanueva de la Ca\~nada, Madrid, Spain
\and
National Astronomical Observatories, Chinese Academy of Sciences,
20A Datun Road, Beijing 100101, China
\and
School of Astronomy and Space Science, University of Chinese Academy of Sciences,
19A Yuquan Road, Beijing 100049, China
\and
School of Astronomy and Space Science, Nanjing University,
Nanjing 210093, China
\and
Key Laboratory of Modern Astronomy and Astrophysics (Nanjing University),
Ministry of Education, Nanjing 210093, China
\and
Dipartimento di Fisica e Astronomia ``Galileo Galilei'',
Universit\`a di Padova, Via F. Marzolo 8, I-35131 Padova, Italy
\and
Mullard Space Science Laboratory, University College London, Holmbury St Mary, Dorking, Surrey RH5 6NT, UK
\and
INAF--Osservatorio Astronomico di Roma, Via Frascati 33,
I-00078 Monte Porzio Catone (RM), Italy
\and
Scuola Universitaria Superiore IUSS Pavia, Piazza della Vittoria 15,
I-27100 Pavia, Italy
\and
INAF--Osservatorio Astronomico di Cagliari, Via della Scienza 5,
I-09047 Selargius (CA), Italy
\and
INAF--Istituto di Radioastronomia, Via P. Gobetti 101,
I-40129 Bologna, Italy
\and
INAF--Istituto di Astrofisica Spaziale e Fisica Cosmica di Milano,
Via Alfonso Corti 12, I-20133 Milano, Italy
\and
Universit\'e Paris-Saclay, Universit\'e Paris Cit\'e, CEA, CNRS,
AIM, F-91191 Gif-sur-Yvette, France
\and
Institute for Frontiers in Astronomy and Astrophysics,
Beijing Normal University, Beijing 102206, China
\and
School of Physics and Astronomy, Beijing Normal University,
Beijing 100875, China
\and
IRAP,
Universit\'e de Toulouse, CNRS, CNES,
9 avenue du Colonel Roche, F-31028 Toulouse, France
\and
Institute of Astronomy, National Tsing Hua University,
Hsinchu 300044, Taiwan
\and
Inter-University Centre for Astronomy and Astrophysics (IUCAA),
Post Bag 4, Ganeshkhind, Pune 411007, India
\and
Department of Physics, The University of Hong Kong,
Pokfulam Road, Hong Kong, China
\and
Max-Planck-Institut f\"{u}r extraterrestrische Physik,
Gie\ss{}enbachstra\ss{}e 1, D-85748 Garching bei M\"{u}nchen, Germany
\and
Universit\'e Paris Cit\'e, CNRS, Astroparticule et Cosmologie,
F-75013 Paris, France
\and
Key Laboratory of Particle Astrophysics, Institute of High Energy Physics,
Chinese Academy of Sciences, Beijing 100049, China
\and
Departamento de F\'isica, Universidad de Santiago de Chile (USACH),
Av. V\'ictor Jara 3493, Estaci\'on Central, Chile
\and
Center for Interdisciplinary Research in Astrophysics and Space Sciences
(CIRAS), Universidad de Santiago de Chile, Chile
\and
The Hong Kong Institute for Astronomy and Astrophysics,
The University of Hong Kong, Pokfulam, Hong Kong, China
\and
Nevada Center for Astrophysics, University of Nevada,
Las Vegas, NV, USA
\and
Department of Physics and Astronomy, University of Nevada,
Las Vegas, NV, USA
\and
Purple Mountain Observatory, Chinese Academy of Sciences,
Nanjing 210023, China
\and
School of Physics, Huazhong University of Science and Technology,
Wuhan 430074, China
\and
Yunnan Observatories, Chinese Academy of Sciences,
Kunming 650011, China
\and
Institute for Gravitational Wave Astronomy, Henan Academy of Sciences,
Zhengzhou 450046, Henan, People's Republic of China
\and
Dipartimento di Fisica, Universit\`a di Trento, via Sommarive 14, I-38123 Povo (TN), Italy
}


\abstract{We report the discovery and early outburst evolution of the new Galactic magnetar \srcfirst, detected by the \ep\ Wide-field X-ray Telescope on 2026 June 28. Follow-up observations with \ep, XMM--Newton, NuSTAR, SVOM, and IXPE revealed coherent X-ray pulsations at $P\simeq6$\,s and an average period derivative of $\dot{P}\simeq2.8\times10^{-12}$\,s\,s$^{-1}$. These values imply a nominal polar dipolar magnetic field of $B_{\rm dip}\simeq2.6\times10^{14}$,G and a characteristic age of $\tau_{\rm c}\simeq34$,kyr, although the structured timing residuals suggest possible torque variability. The rms pulsed fraction is approximately 20--25\% below $\sim7$\,keV and decreases at higher energies. The broadband spectra require multiple thermal components and a hard power-law tail. Adopting a distance of 3.3\,kpc, the 0.5--30,keV luminosity declined from $1.2\times10^{35}$ to $5.7\times10^{34}$,erg,s$^{-1}$ during the first month, accompanied by a decrease in the inferred thermal-emitting areas. At this distance, $b=-4.56^\circ$ corresponds to a height of approximately 0.26\,kpc below the Galactic plane, which is difficult to reconcile with the nominal characteristic age under a simple midplane-birth scenario even for an extreme proper motion velocity. Short X-ray bursts independently confirm the magnetar nature of the source. No near-infrared or radio counterpart were detected by GTC, Medicina or FAST, respectively, down to $K_{\rm s}>21$,mag and $S_{1.25\,{\rm GHz}}<2.3,\mu$Jy. These observations demonstrate the potential of \ep's wide-field monitoring to uncover previously quiescent Galactic magnetars and follow their outbursts from their earliest observed stages.
}


\maketitle
\nolinenumbers

\section{Introduction}
\label{sec:introduction}

Magnetars are isolated neutron stars whose emission is powered predominantly by the decay and reconfiguration of ultra-strong magnetic fields \citep{duncan92, thompson00}. They are characterised by spin periods of a few seconds, large spin-down rates, luminous and variable X-ray emission, and, most distinctively, short energetic X-ray and soft gamma-ray bursts (e.g., \citealt{esposito21,rea26}). Most known magnetars spend extended intervals in comparatively faint states and are discovered when they undergo large-amplitude X-ray outbursts, during which their luminosity can increase by several orders of magnitude \citep{cotizelati18}. Detecting such events from their onset is particularly valuable for probing the mechanisms responsible for magnetar activation and for following the accompanying evolution of their thermal emission, magnetospheric currents, and rotational properties. 
The wide-field soft X-ray monitoring capability of \ep\ (EP; \citealt{Yuan22,Yuan2025}) provides a new opportunity to discover previously dormant magnetars at the onset of their outbursts. On 2026 June 28, the Wide-field X-ray Telescope (WXT) aboard EP discovered a previously unknown transient \srcfirst\ (EP260628c) at Galactic coordinates $l=103.648^\circ$, $b=-4.576^\circ$, with an unabsorbed 0.5--4\,keV flux of $\sim1.3\times10^{-10}$\,\flux\ and a possible modulation in the X-ray emission \citep{Yang2026ATel17859}. Follow-up observations with the EP Follow-up X-ray Telescope (FXT) confirmed a bright X-ray source and revealed the exact spin period \cite[$P\approx 6\, \mathrm{s}$;][]{Rea2026ATel17870} which, together with the discovery of short bursts by SRG/ART-XC \citep{Molkov2026ATel17908} and SVOM/MXT \citep{Feng2026GCN45270}, confirmed its magnetar nature. Furthermore, IXPE detected a significant $\sim7$\% polarization degree during the outburst (Taverna et al. 2026, submitted).
In this paper, we present the discovery of \srcfirst\ (hereafter \src), focusing on its X-ray early-time evolution, timing properties, and broadband spectral evolution as well as infrared and radio multi-band properties using data from EP, \nustar, \xmm, SVOM, IXPE, GTC, Medicina, and FAST (see \S\ref{sec:data reduction}). The results of our multi-band analysis are presented in \S\ref{sec:results}, while in \S\ref{sec:discussion} we discuss our results and future prospects.

\section{Observations and data reduction} \label{sec:data reduction}

\subsection{X-ray observations}
\subsubsection{\ep} \label{subsec:ep}

The \textit{Einstein Probe} (EP) mission carries two complementary X-ray instruments: the WXT \citep{cheng2026} 
and the FXT \citep{chen20}.
WXT employs lobster-eye micro-pore optics and covers a field of view of $\sim3600\,{\rm deg}^2$ in the 0.5--4\,keV energy band, reaching a sensitivity of $(2-3)\times10^{-11}\,\flux$; its energy resolution is $\sim170$\,eV at 1\,keV and its angular resolution is $\approx5\arcmin$ (full width at half maximum; FWHM). The FXT is sensitive over 0.3--10\,keV with a typical localisation accuracy of $\sim10\arcsec$ (90\% confidence), an energy resolution of $\sim120$\,eV at 1.25\,keV, and a sensitivity of $\sim10^{-14}\,\flux$. It comprises two co-aligned modules, FXT-A and FXT-B, which can operate in Full Frame (FF), Partial Window (PW), or Timing (TM) modes, providing time resolutions of 50\,ms, 2\,ms, and 23.6\,$\mu$s, respectively.
The source \src\ was first detected by WXT on 2026 June 28 at 16:22:49 UTC and was observed 23 times by FXT between the discovery epoch and 2026 August 22, accumulating a total exposure of $\sim112$\,ks (see Table\,\ref{tab:log}).  

We processed the data using \texttt{fxtchain} within the FXT Data Analysis Software Package (\texttt{FXTDAS}; v1.30), following the standard calibration and screening procedures described in the FXT User Guide.\footnote{\url{https://epfxt.ihep.ac.cn/analysis}} Additional filtering and event selection were carried out with \texttt{Xselect}. For FF and PW observations, source events were extracted from a circular region of radius $60\arcsec$, while the background was estimated from an annulus with inner and outer radii of $120\arcsec$ and $240\arcsec$, respectively; for observations affected by pile-up, the central $5\arcsec$ of the source extraction region was excluded. For TM observations, source and background events were instead selected using $180\arcsec\times60\arcsec$ rectangular regions oriented according to the spacecraft roll angle, with the background region displaced by $360\arcsec$ from the source position. Individual FXT snapshots had exposures ranging between 2.1 and 6.3\,ks.

\subsubsection{\nustar}\label{subsec:nustar}
\nustar\ \citep{Harrison2013} observed \src\ four times between its discovery and 2026 August 22, for a total on-source exposure time of 90.5\,ks and 89.5\,ks for the focal plane modules A and B (FPMA and FPMB hereafter), respectively. We processed the event lists and filtered out passages of the satellite through the South Atlantic Anomaly using the tool \texttt{NUPIPELINE}. Both source and background counts were accumulated within a circular region of radius 90\arcsec. We then applied the script \texttt{NUPRODUCTS} to extract light curves and spectra, and generate response files for both FPMs.

\subsubsection{\xmm}\label{subsec:xmm}
\src\ was observed with the European Photon Imaging Cameras (EPIC) on board \xmm\ on 2026 July 3--4, for an exposure time of $\sim$12\,ks. The EPIC-pn \citep{struder01} was set in Small Window mode (SW; timing resolution of 5.7\,ms), while the EPIC-MOS1 and EPIC-MOS2 cameras \citep{turner01} were operating in SW and Full Frame (FF) mode (timing resolution of 0.3\,s and 2.6\,s), respectively. 
We consider only the data acquired with the EPIC-pn for the spectral analysis, which provides the data set with the highest counting statistics owing to its larger effective area compared to the EPIC-MOS cameras and free of pile-up. For the timing analysis, data from EPIC-MOS1 were also included; EPIC-MOS2 in FF mode does not offer high enough timing resolution to resolve the two peaks of the pulse profile (see Sec.\,\ref{subsec:timing}). 
 
Raw data were reprocessed and analysed with standard prescriptions. We cleaned the observation from periods of high background activity resulting in a net exposure of 2.9\,ks and 1.2\,ks for the EPIC-pn and EPIC-MOS1, respectively. For both datasets, the source events were selected from a circle with a radius of 30\arcsec\ and the background counts were accumulated from a nearby circle of the same size. The response matrices and ancillary files were generated through the \texttt{RMFGEN} and \texttt{ARFGEN} tools, respectively. 

\subsubsection{SVOM}\label{subsec:svom}

The Microchannel X-ray telescope \citep[MXT; ][]{mxt} on board the Space Variable astronomical Object Monitor \citep[SVOM; ][]{svom} has observed the position of \src\ three times in the framework of our approved
ToO programme on magnetars. 

The first observation was conducted on July 5$^{th}$ starting at 15:25:07 UTC and offered an exposure of 3.47 ks. The second observation started at 17:32:09 UTC on July 7$^{th}$ and cumulated 2.6 ks of exposure time, while the last one started on July 8$^{th}$ at 14:32:57 UTC for an exposure of 3.5 ks. The MXT data were reduced using version 1.16.4 of the MXT pipeline \citep{mxtpipeline}, selecting the grades 0--12 in the 0.2 -- 10 keV energy range. Source events were collected within the standard cross-shaped region centred on the source detection position, and the background was cumulated of the rest of the detector. Event files have been Earth-barycentered, and light curves and OGIP standard\footnote{\url{https://heasarc.gsfc.nasa.gov/docs/heasarc/ofwg/docs/spectra/ogip_92_007/ogip_92_007.html}} spectra have been extracted.

\subsubsection{\swift}\label{subsec:swift}
The sky region containing \src\ was serendipitously observed twice with the X-Ray Telescope (XRT; \citealt{2005SSRv..120..165B}) aboard the \textit{Neil Gehrels Swift Observatory} \citep{2004ApJ...611.1005G} in photon-counting mode, on 2017 February 28 for 52\,s (ObsID 07006980001) and 2017 March 4 for 484\,s (ObsID 07006980002). The data were processed following standard procedures. We selected events in the 0.3--10\,keV band with grades 0--12. Source events were extracted within a circular region of radius $47.2\arcsec$, centred on the source position and enclosing about 90\% of the point-spread function at 1.5~keV. The background was estimated from an annulus with inner and outer radius of 142 and 260\arcsec. We then stacked the two event lists and summed the corresponding exposure maps. No events were detected within the source aperture of the stacked image. We estimated a 3-$\sigma$ upper limit on the 0.3--10~keV net count rate of 0.014\,counts\,s$^{-1}$ using the \texttt{SOSTA} task in \texttt{XIMAGE}.

\begin{figure}
\centering
\includegraphics[width=0.48\textwidth]{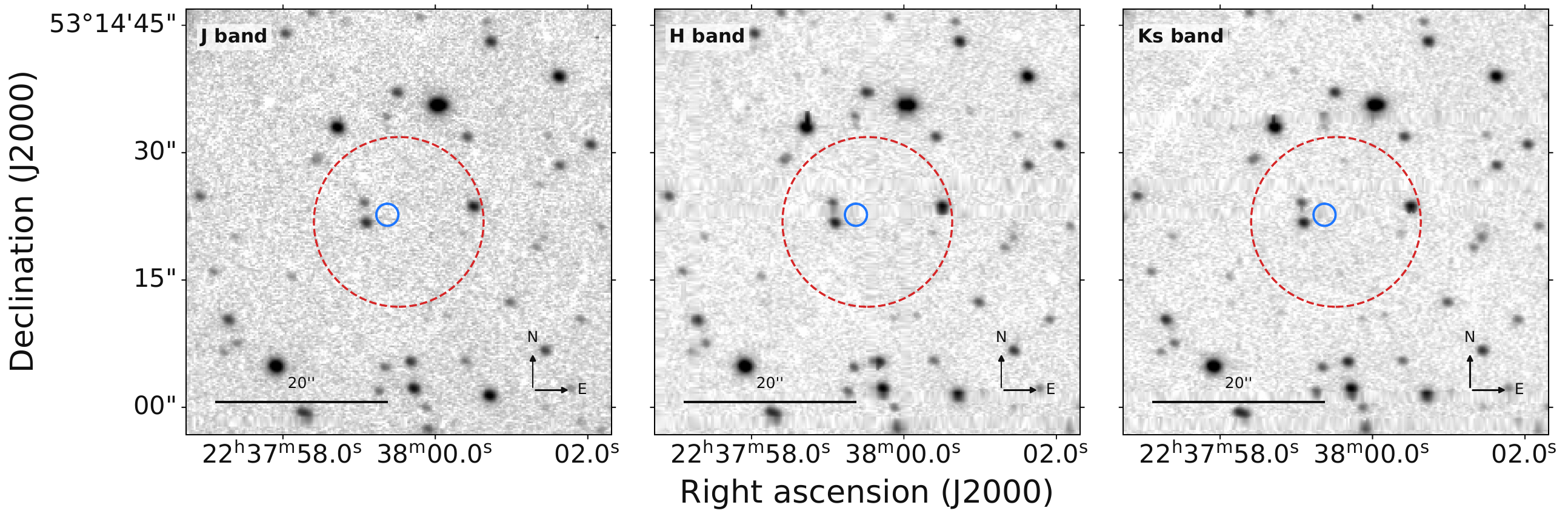}
\caption{GTC/EMIR near-infrared stacked images of the field of \src\ in the $J$, $H$, and $K_{\rm s}$ bands. The dashed red circle indicates the $10\arcsec$-radius 90\% confidence EP/FXT localisation, while the solid cyan circle indicates the $1.3\arcsec$-radius \xmm\ localisation. North is up and east is to the right.}
\label{fig:nir_opt}
\end{figure}

\subsection{Infrared observations}
\label{subsec:gtc}
We obtained near-infrared imaging of the field of \src\ with the Espectr\'ografo Multiobjeto Infra-Rojo (EMIR; \citealt{Garzon2022}) mounted on the 10.4-m Gran Telescopio Canarias (GTC) on La Palma, Canary Islands. The observations were carried out on 2026 July 4 from 03:49:20 to 04:46:20 UTC. Dithered images were acquired in the $J$, $H$, and $K_{\rm s}$ filters, with total integration times of 195, 325, and 1200\,s, respectively. The final stacked images had stellar FWHMs of $0\farcs84$, $0\farcs78$, and $0\farcs77$ in $J$, $H$, and $K_{\rm s}$, respectively.

The data were reduced with the standard EMIR pipeline, including bad-pixel correction, flat-fielding, sky subtraction, astrometric reprojection, and sigma-clipped image combination. The final images were astrometrically registered to Gaia DR3 \citep{GaiaCollaboration2023}, after propagating Gaia proper motions to the epoch of the observations. The median radial astrometric residuals were $0\farcs19$, $0\farcs20$, and $0\farcs21$ in $J$, $H$, and $K_{\rm s}$, respectively, while the median relative astrometric offsets between the registered images were $0\farcs05$ ($J-H$), $0\farcs09$ ($J-K_{\rm s}$), and $0\farcs05$ ($H-K_{\rm s}$). Figure~\ref{fig:nir_opt} shows the final $JHK_{\rm s}$ stacks.

\subsection{Radio observations}

\label{subsec:radio}
We observed \src\ with the 32-m Medicina radio telescope on 2026 July 3--4 and July 10, for 3.03 and 3.44\,hr of on-source time, respectively. Data were acquired with the Digital Base Band Converter, recorded in VDIF format and converted to a standard filterbank format  with 8-bit sampling. After visual inspection the band was reduced to a clean 64\,MHz centred at 1.4\,GHz, divided into 256 channels of 250\,kHz each. The L-band receiver has $T_{\rm sys}=55$\,K and a gain of 0.12\,K\,Jy$^{-1}$ ($\mathrm{SEFD}\simeq460$\,Jy). Both epochs are largely simultaneous with pointed X-ray observations.


We finally conducted two observations of \src\ using the Five-hundred-meter Aperture Spherical Telescope (FAST) \citep{Nan_2013}. The observations were carried out on 4 July 2026 (20:41:59--21:41:59 UTC) and 2 August 2026 (18:40:59--19:40:59 UTC). A 1-minute noise-diode calibration at a frequency of 4.967053731282552\,Hz was performed at both the beginning and end of each observation, leaving 58 minutes of on-source integration time per epoch. The observations were carried out in L-band tracking mode with the FAST ROACH backend, which recorded full Stokes polarization \citep{2020RAA....20...64J}. The data were centered at 1250\,MHz with a total bandwidth of 400\,MHz covering 1.05--1.45\,GHz, divided into 4096 frequency channels, with a sampling interval of 49.152\,\(\mu\mathrm{s}\).

\section{Data analysis and results} \label{sec:results}
In the following, we derive all quantities assuming a distance of 3.3\,kpc (see\,Sec.\,\ref{sec:distance}) and all uncertainties are quoted at 1$\sigma$ confidence level (c.l.) unless otherwise stated.

\subsection{X-ray localisation}
\label{sec:xray-localization}

We refined the X-ray position of \src\ using the \xmm/EPIC data. We ran the SAS source-detection tool \texttt{edetect}$_-$\texttt{chain} and obtained the final source parameters from a simultaneous maximum-likelihood fit  to the three EPIC cameras. \src\ is detected at the best-fitting position R.A. $=22^{\rm h}37^{\rm m}59.370^{\rm s}$ and decl. $=+53^\circ14\arcmin22.68\arcsec$ (J2000.0), which corresponds to l=103.6533$^{\deg}$ and b=--4.562$^{\deg}$. The formal statistical uncertainty returned by \texttt{emldetect} is $0.01\arcsec$, whereas the systematic uncertainty estimated by \texttt{catcorr} is $1.3\arcsec$. We therefore adopt a $1\sigma$ positional uncertainty of $1.3\arcsec$.

\begin{table}[ht]
\centering
\caption{Timing solution for \src.}
\label{tab:timing_solution}
\renewcommand{\arraystretch}{1.15}
\begin{tabular}{lc}
\hline
\hline
Parameter & Value \\
\hline
Epoch (BMJD$_{\rm TDB}$) & 61220.0 \\
Validity range (BMJD$_{\rm TDB}$) & 61220.558--61274.369 \\
$F_0$ (Hz) & $0.166781381(2)$ \\
$F_1$ (Hz\,s$^{-1}$) & $(-7.79\pm0.08)\times10^{-14}$ \\
$P$ (s) & $5.99587314(6)$ \\
$\dot P$ (s\,s$^{-1}$) & $(2.80\pm0.03)\times10^{-12}$ \\
Number of TOAs & 61 \\
$\chi^2/{\rm dof}$ & $96.74/54$ \\
Weighted residual rms (ms) & 25.4 \\
\hline
$B_{\rm dip}$ (G) & $2.6\times10^{14}$ \\
$\tau_{\rm c}$ (kyr) & 34 \\
$\dot E$ (erg\,s$^{-1}$) & $5.1\times10^{32}$ \\
\hline
\end{tabular}
\tablefoot{Derived quantities assume the standard vacuum-dipole estimates for
the surface magnetic field at the poles $B_{\rm dip}\simeq6.4\times10^{19}(P\dot P)^{1/2}$~G,
the characteristic age $\tau_{\rm c}\simeq P/(2\dot P)$, and
the rotational energy loss rate $\dot E \simeq 4\pi^2 I\dot P/P^3$ 
with an assumed moment of inertia of $I=10^{45}$~g~cm$^2$ (see also Fig.\,\ref{fig:ppdot}) }
\end{table}
 
\begin{figure}
\centering
\includegraphics[width=\columnwidth]{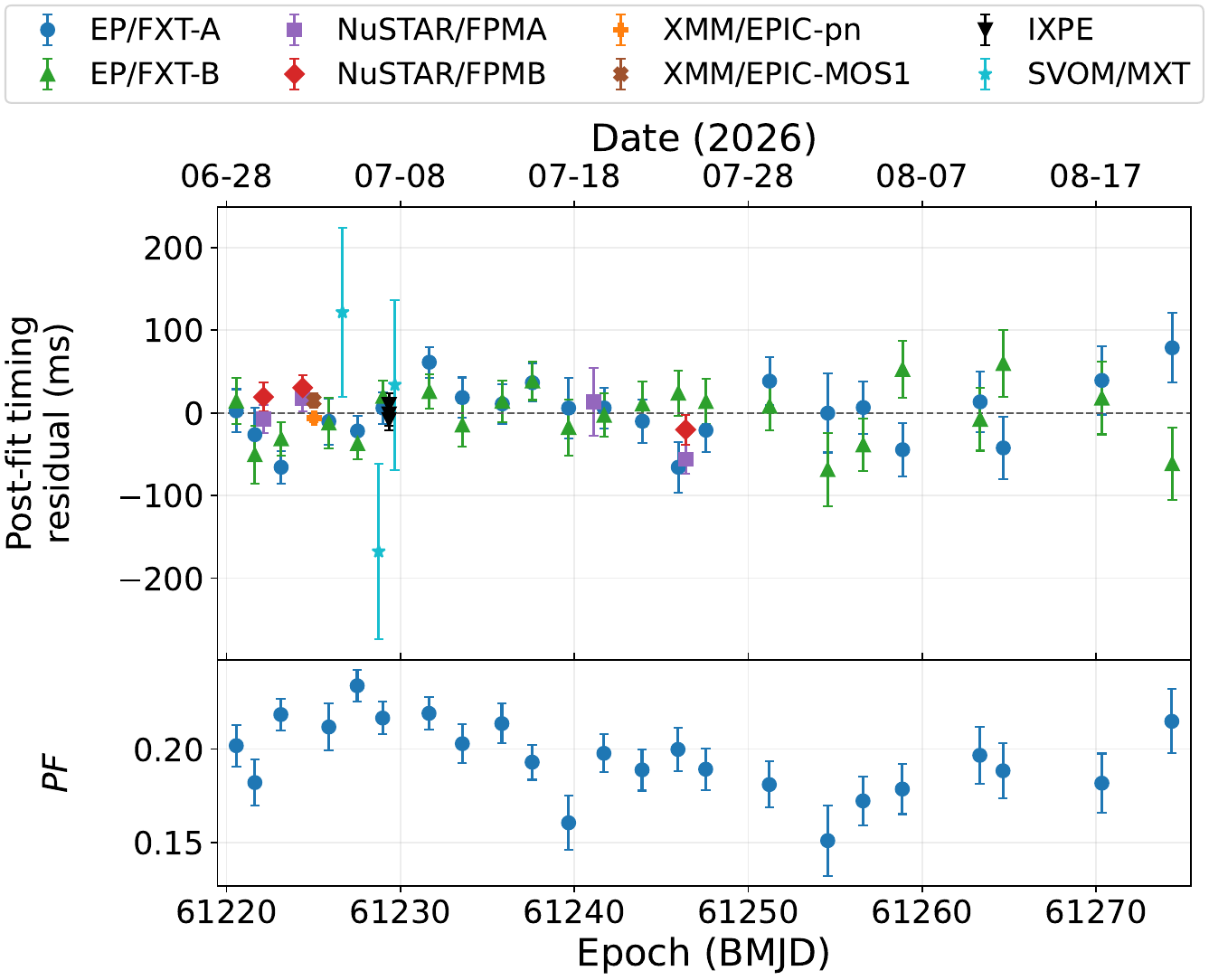}
\caption{\emph{Top}: post-fit timing residuals for the phase-connected timing solution reported in Table~\ref{tab:timing_solution}, including the EP/FXT, IXPE, \nustar\, \xmm\ EPIC-pn and MOS1 and SVOM/MXT Times of Arrival (TOAs). \emph{Bottom}: background-subtracted rms pulsed fraction, $PF$, as a function of time for EP/FXT-A.}
\label{fig:timing_residuals}
\end{figure}

\begin{figure}
\centering
\includegraphics[width=0.48\textwidth]{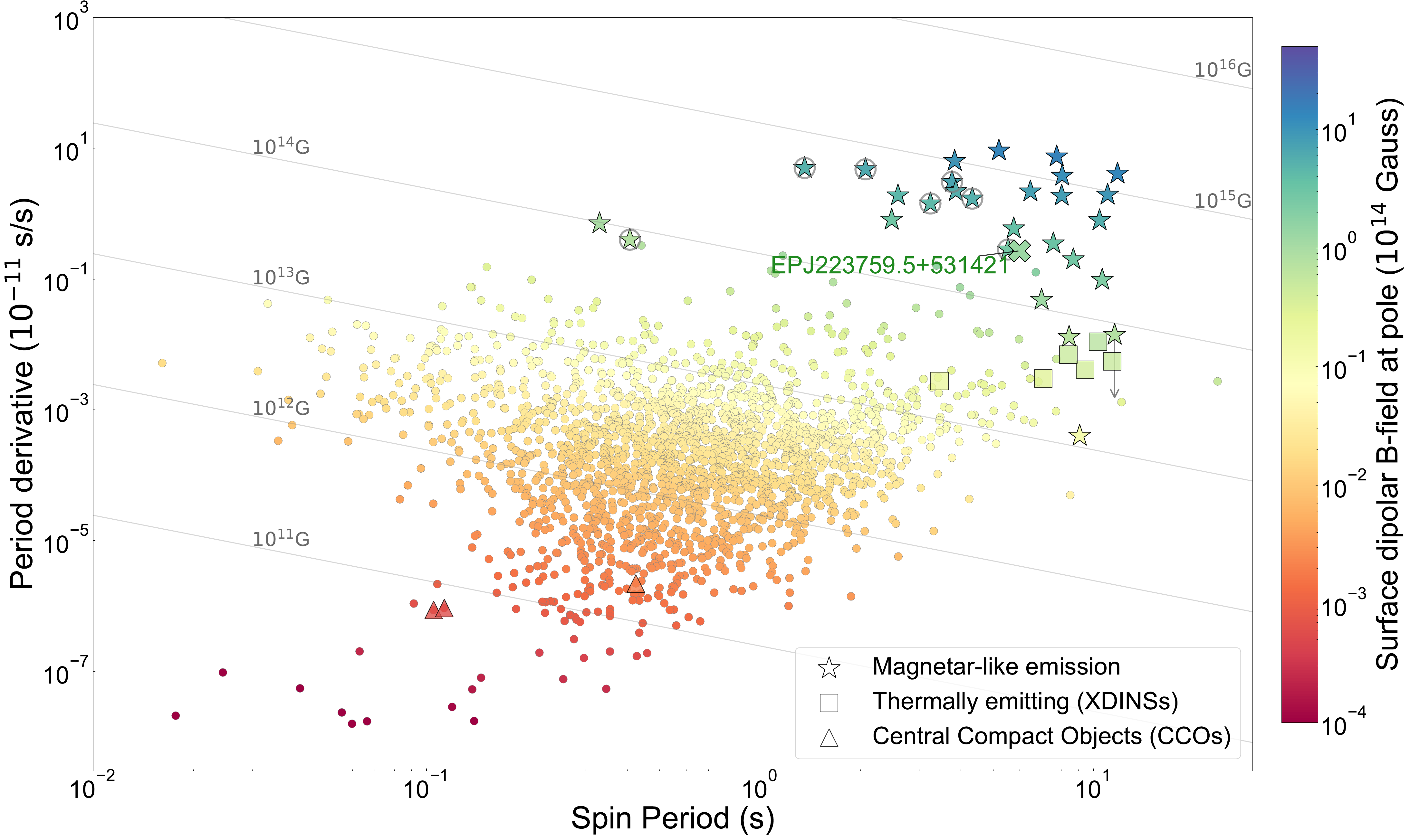}
\caption{Spin Period versus Period derivative for different isolated neutron star classes: \src\, is labelled as a cross. The colour bar indicates the surface magnetic field at pole as measured under the dipolar spin-down assumption.}
\label{fig:ppdot}
\end{figure}

\subsection{X-ray timing analysis}
\label{subsec:timing}

We referred the photon arrival times to the Solar System barycentre using our best \xmm\ position (see Section\,\ref{sec:xray-localization}) and ephemeris DE430 \citep{Folkner2014}. 

Pulsations were first detected in the event lists from the first EP/FXT observation using a Leahy-normalized power spectrum of a light curve sampled at 0.1\,s \citep{leahy83_ray}. Peaks were detected at 0.166784\,Hz and its second harmonic, corresponding to $P\simeq5.9958$\,s, with a trial-corrected significance of $\simeq$14$\sigma$. This detection provided the provisional ephemeris used to assign pulse phases and to construct initial empirical Fourier pulse templates for each dataset for the subsequent time-of-arrival (TOA) analysis.
For each dataset, the template phase offset was determined through an unbinned likelihood fit and converted into a TOA. We then fitted the TOAs using the phase-connected rotational model $\Phi(t)=\Phi_0+F_0(t-t_0)+\frac{1}{2}F_1(t-t_0)^2+...$, where $F_0$ and $F_1$ are the spin frequency and its first derivative. We tested successively higher-order frequency derivatives, retaining an additional term only if it improved the fit by $\Delta\chi^2>9$, corresponding approximately to a $3\sigma$ improvement for one additional degree of freedom (dof).
For this analysis, we also included IXPE data presented by Taverna et al. (submitted).
The final dataset comprised 61 TOAs over the time interval 61220.558--61274.369 Barycentric Modified Julian Date (BMJD), BMJD$_{\rm TDB}$.

\begin{figure*}
\centering
\includegraphics[width=0.75\columnwidth]{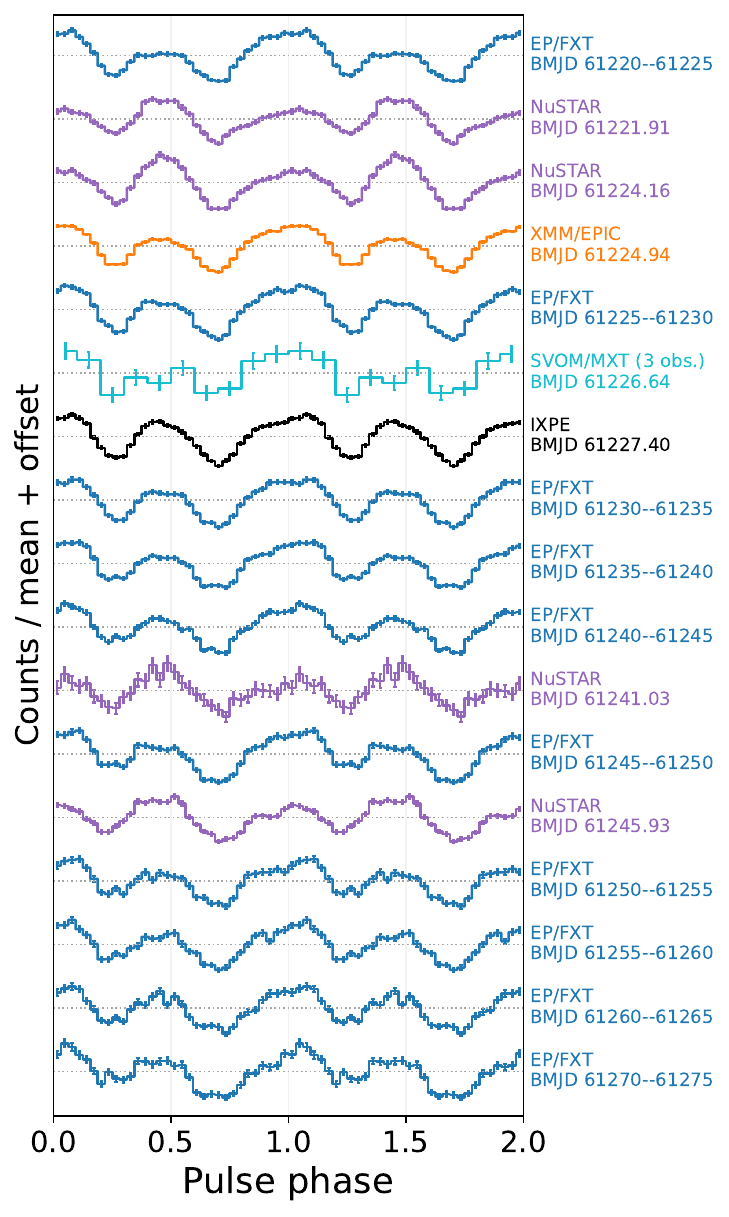}
\includegraphics[width=\columnwidth]{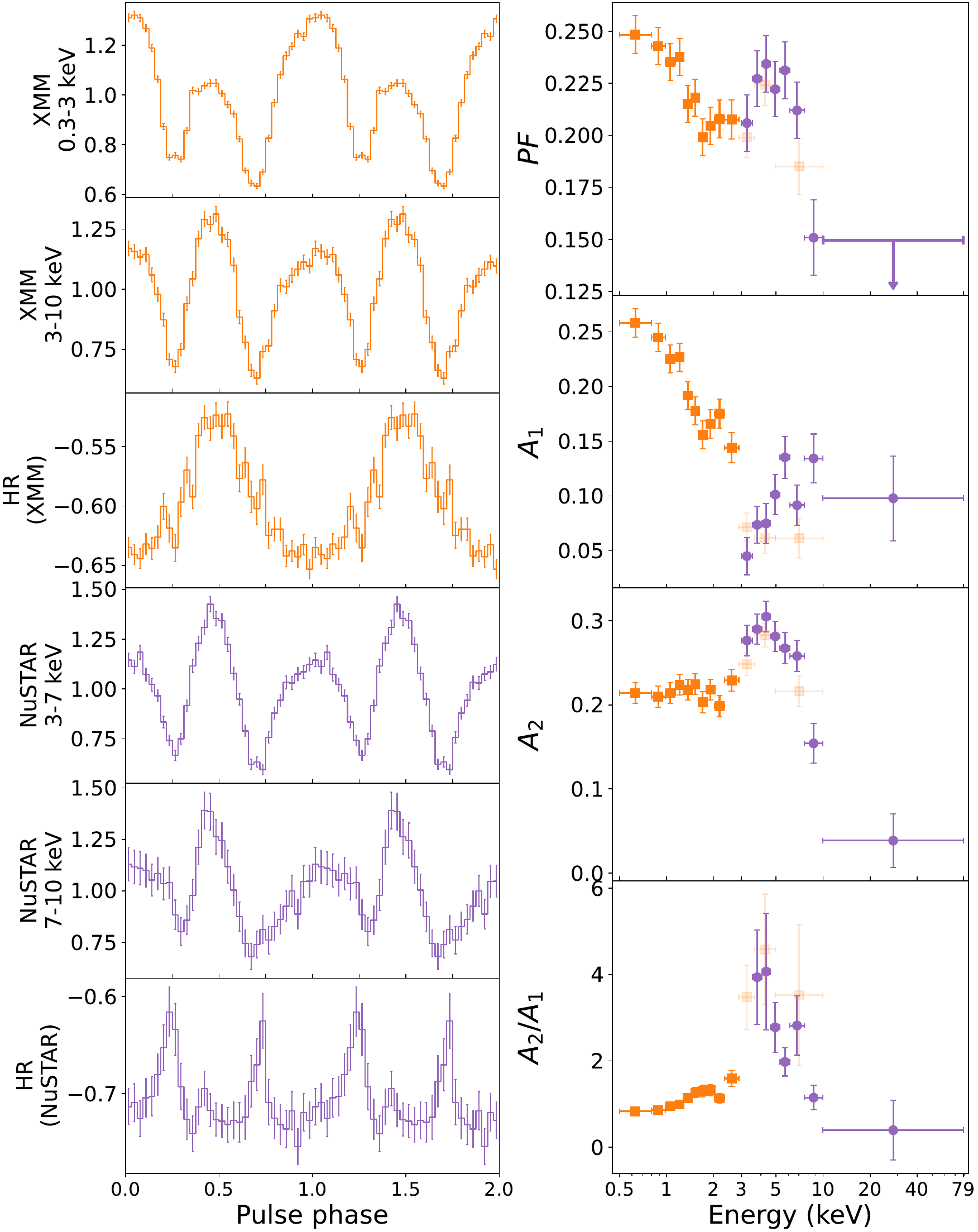}
\caption{{\em Left.} Background-subtracted pulse profiles of \src{} folded using our timing solution. The profiles are plotted over two pulse cycles and shifted vertically for display purposes. The 0.5--10\,keV EP profiles combine the FXT-A and FXT-B event lists and are grouped into five-day intervals. The NuSTAR profiles combine the FPMA and FPMB event lists. The IXPE profile combines the event lists from its three detectors, while the \xmm\ profile combines EPIC-pn and EPIC-MOS1 data. The \textit{SVOM}/MXT profile combines the three observations in the 0.5--5\,keV band. {\em Middle.} Energy dependence of the pulse profile of \src\ from quasi-simultaneous \xmm\ (orange) and \nustar\ (purple) observations. The background-subtracted pulse profiles are normalised to their mean net count rates; the corresponding hardness ratios are defined as ${\rm HR}=(H-S)/(H+S)$, where $H$ and $S$ are the background-subtracted hard- and soft-band count rates, respectively. Two phase cycles are shown for clarity. {\em Right.} Background-subtracted rms pulsed fraction, $PF$, Fourier amplitudes, and their ratio as a function of photon energy. The \xmm\ measurements above 3\,keV are plotted in a fainter orange to emphasize the overlapping \nustar\ measurements, which provide tighter constraints in this energy range.}
\label{fig:folded_profiles}
\label{fig:xmm_nustar_phase_energy_harmonics}
\end{figure*}

The resulting timing solution comprises only the $F_0$ and $F_1$ terms and is given in Table~\ref{tab:timing_solution}. At a reference epoch of 61220.0 BMJD$_{\rm TDB}$, we obtain $P=5.99587314(6)\ {\rm s}$ and $\dot P=(2.80\pm0.03)\times10^{-12}\ {\rm s\,s^{-1}}$ (see also Fig.\,\ref{fig:ppdot}). The large reduced $\chi^2$ and structured post-fit residuals indicate additional timing noise and/or torque variability; accordingly, we interpret the measured $\dot P$ as the average spin-down over the observed interval rather than as a precise measurement of the long-term secular value. Figure~\ref{fig:timing_residuals} shows the post-fit TOA residuals, while Fig.~\ref{fig:folded_profiles} shows the pulse profiles folded using this timing solution.

\subsubsection{Pulse profile evolution}
The pulse profile remains broadly stable throughout the monitoring campaign, with no compelling evidence for a change in the pulse morphology.
We define the rms pulsed fraction following the fractional-rms formalism of \citet{Vaughan2003}, that is, $PF=\sqrt{\left(S^2-\overline{\sigma_{\rm err}^2}\right)}/\bar{x}$, where $S^2$ is the sample variance of the light curve, $\overline{\sigma_{\rm err}^2}$ is the mean squared measurement uncertainty, and $\bar{x}$ is the mean count rate. The $PF$ over the 0.5--10\,keV energy interval ranges from $\approx$0.16 to 0.23 throughout the observations, with no evidence for a monotonic evolution over the time span covered by the current data (Fig.~\ref{fig:timing_residuals}, bottom). 

To investigate the energy dependence of the pulse shape, we extracted background-subtracted folded profiles and hardness ratios for both \xmm\ and the second \nustar\ observation, which are quasi-simultaneous (Fig.~\ref{fig:xmm_nustar_phase_energy_harmonics}). The \xmm\ profiles combine EPIC-pn and EPIC-MOS1 data in the 0.3--3 and 3--10\,keV bands. The pulse shape changes substantially between these bands: the relative strength of the two pulse components is reversed. For \nustar\, we combined FPMA and FPMB data and used 3--7 and 7--10\,keV bands. The corresponding profile and HR variations show that the pulse shape continues to evolve above the \xmm\ bandpass. 

We also measured the background-subtracted rms pulsed fraction and Fourier harmonic content as a function of energy (Fig.~\ref{fig:xmm_nustar_phase_energy_harmonics}). The pulsed fraction remains at $\approx$0.20--0.25 below $\sim$7\,keV and decreases to $\approx$0.15 in the 7.5-–10\,keV band. At higher energies, we obtain a 95\% upper limit of $PF<0.15$ in the 10-–79\,keV band.
The harmonic decomposition indicates a substantial redistribution of pulsed power with energy: in the \xmm\ data, the amplitude of the fundamental  component $A_1$ decreases towards higher energies, whereas that of the second harmonic component $A_2$ varies more moderately. Hence, their ratio is enhanced in the 3--7\,keV \nustar\ bands. At higher energies, both $A_2$ and $A_2/A_1$ decrease. Overall, the energy-dependent pulse morphology is driven by changing relative contributions of the fundamental and second harmonic components.

\subsection{X-ray spectral analysis} \label{subsec:spec_phaseave}
The \xmm/EPIC-pn and \nustar\ background-subtracted spectra were grouped so as to have at least 50 counts per energy bin, while the EP/FXT spectra were binned to guarantee at least 25 counts per energy bin. We used only the FXT data collected with imaging modes (PW and FF) for the spectral analysis, since TM spectra suffer from higher systematic uncertainties.
The spectral analysis was performed with the \textsc{xspec} package \citep{arnaud96}, adopting the \textsc{Tbabs} model to describe the interstellar absorption with chemical abundances from \citealt{Wilms2000} and photoionization cross-section from \citealt{verner96}.

\begin{figure}
\centering
\includegraphics[width=\columnwidth]{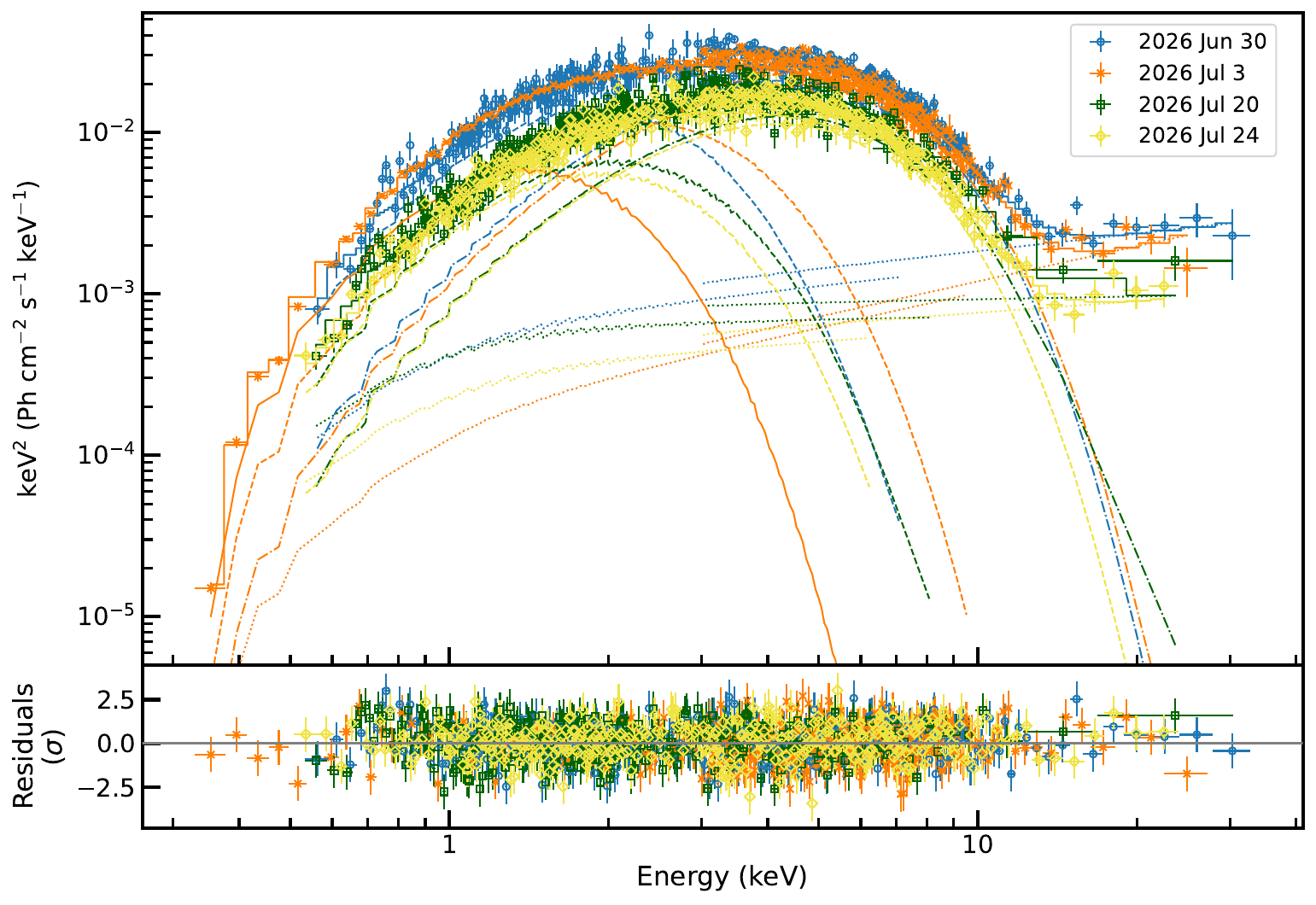}
\caption{Broadband $E^2f(E)$ unfolded spectra for the (quasi-)simultaneous EP, \xmm\ and \nustar\ observations performed on 2026 June 30 (blue), July 3 (orange), July 20 (green), July 24 (yellow). The best-fitting models are plotted with a solid line, together with the different spectral components.
Post-fit residuals in units of standard deviations are shown in the bottom panel.}
\label{fig:spec_pha_ave}
\end{figure}

We started the fitting procedure by modelling the broadband spectra extracted from the (quasi-)simultaneous observations acquired with \xmm, \nustar\ and EP for the epochs 2026 June 30, July 3, 20 and 24. We fitted the spectra jointly, allowing all the parameters to vary between the different epochs except for the hydrogen column density \nh, which was forced to maintain the same value in all datasets.
A renormalisation factor was included to account for cross-calibration uncertainties among the different instruments and was fixed at 1 for \xmm/EPIC-pn. A model composed of two absorbed blackbodies did not provide a good fit, and revealed structured residuals above $\sim$10\,keV. 
The addition of a power law improved the fit for the nearly simultaneous EP and \nustar\ spectra (i.e., epochs 2026 June 30, July 20 and 24; $F$-test probability $<10^{-10}$, corresponding to $>7\sigma$). However, to obtain a satisfactory description of the \xmm/EPIC-pn+\nustar\ spectra (i.e., epoch 2026 July 3), we needed to include an extra blackbody component ($F$-test probability $\sim 6\times10^{-6}$, equal to 4.7$\sigma$). 
The fit yielded \nh=(2.4$\pm$0.2)$\times$10$^{21}$\,cm$^{-2}$ ($\chi^2$=1890.82 for 1854 dof). The best-fitting values are listed in Table\,\ref{tab:broadband_fit} and Fig.\,\ref{fig:spec_pha_ave} shows the broadband spectra. The 0.5--30\,keV luminosity decreased by a factor of $\sim$2 between 2026 June 30 (two days after the discovery) and July 24, going from (1.2$\pm$0.1)$\times10^{35}$\,\lum\ to (5.7$\pm$0.1)$\times10^{34}$\,\lum. We did not detect a clear time evolution of the spectral parameters, except for the radii of the warm and hot emitting blackbody regions that decreased in time by a factor of $\sim$1.6 and 1.3, respectively.

\begin{figure}
\centering
\includegraphics[width=0.85\columnwidth]{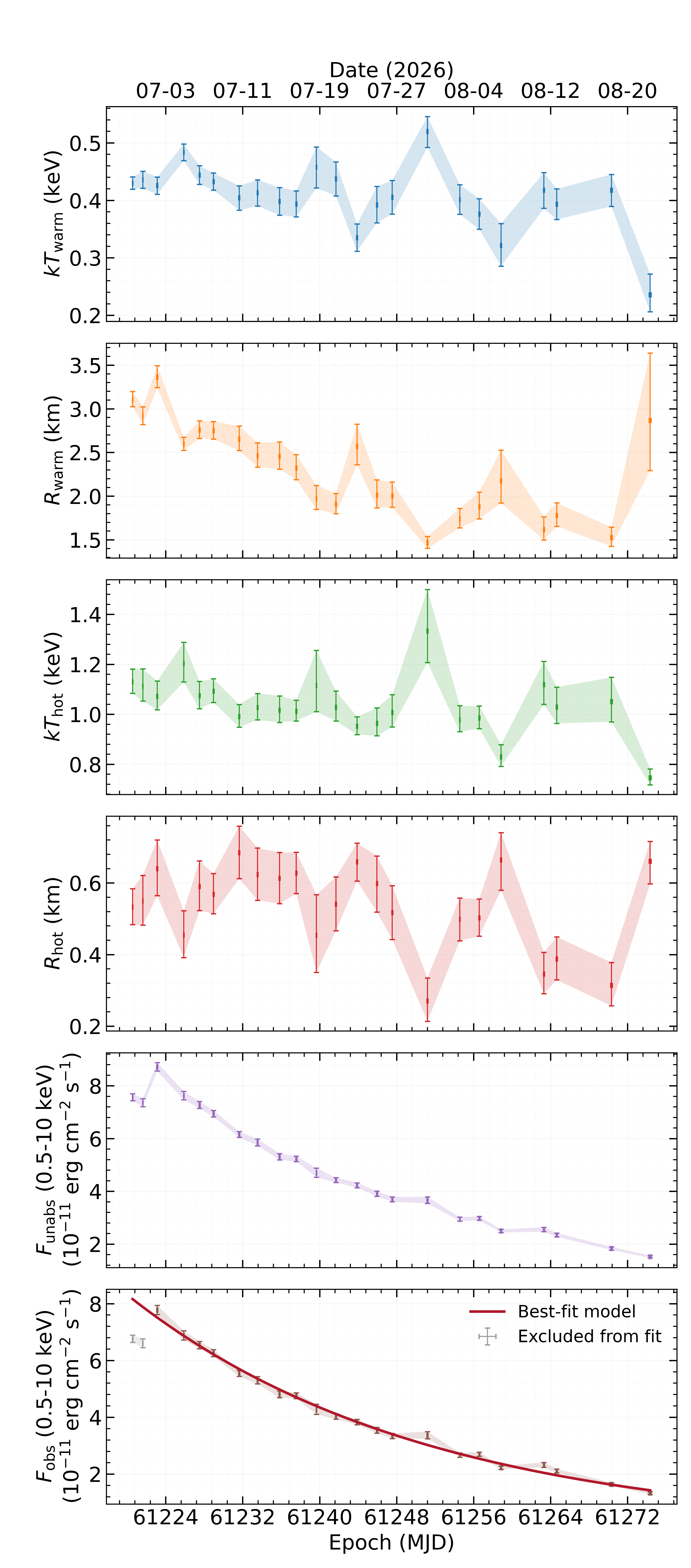}
\caption{The evolution of spectral parameters derived from the analysis of EP/FXT spectra. From top to bottom, the parameters are, respectively, temperature ($kT_{\rm warm}$) and radius ($R_{\rm warm}$) of the warm black body component, temperature ($kT_{\rm hot}$) and radius ($R_{\rm hot}$) of the hot black body component, the unabsorbed flux ($F_{\rm unabs}$), and the observed flux ($F_{\rm obs}$) together with an exponential fit to the observed decay of the FXT flux (see the text for the best-fit parameters of the exponential model). The model adopted in spectral fitting is $\texttt{TBabs}\times(\texttt{bbodyrad}+\texttt{bbodyrad})$. The hydrogen column density is fixed to the value derived from the broadband spectral analysis, $N_{\rm H}=2.4\times10^{21}\,{\rm cm^{-2}}$. The radii are calculated considering a distance of 3.3\,kpc.}
\label{fig:ep_spec_evolution}
\end{figure}

To characterise the long-term evolution of the outburst, each FXT spectrum was fitted with an absorbed double-blackbody model, fixing the hydrogen column density to the value inferred from the broadband spectral analysis ($N_{\rm H}=2.4\times10^{21}\,{\rm cm^{-2}}$) and allowing all other parameters to vary. Satisfactory fitting results (with $\chi^2/{\rm dof}<1.5$) were derived for all observations. The evolution of the 0.5--10\,keV flux, together with that of other parameters, is shown in Figure~\ref{fig:ep_spec_evolution}. The best-fitting results are also listed in Table\,\ref{tab:fxt_spectral_fit}.

To characterise the long-term decline of the absorbed X-ray flux, we fitted the 0.5--10\,keV flux evolution with a single-exponential model,
\begin{equation}
F_{\rm obs}(t)=A\exp\left(-\frac{t-t_0}{\tau}\right),
\end{equation}
where $A$ is the normalisation and $\tau$ is the $e$-folding timescale. The reference epoch $t_0$ was fixed at the time of the first FXT observation. The first two flux measurements were made in the flux-rising phase, and were therefore excluded from the decay fit. We fitted the remaining measurements by minimising $\chi^2$ using a weighted nonlinear least-squares method. The best-fitting model gives a normalisation of $A=(8.16\pm0.05)\times10^{-11}$\,\flux\ and an e-folding timescale of $\tau=30.9\pm0.2$\,d. The fit yields $\chi^2/{\rm dof}=132/19$, corresponding to a reduced chi-square of $\chi^2_\nu\approx7.0$. The best-fitting model is shown in the bottom panel of Figure~\ref{fig:ep_spec_evolution}.

\subsection{Pulse-phase spectroscopy analysis} \label{subsec:spec_pps}

\begin{figure}
\centering
\includegraphics[width=\columnwidth]{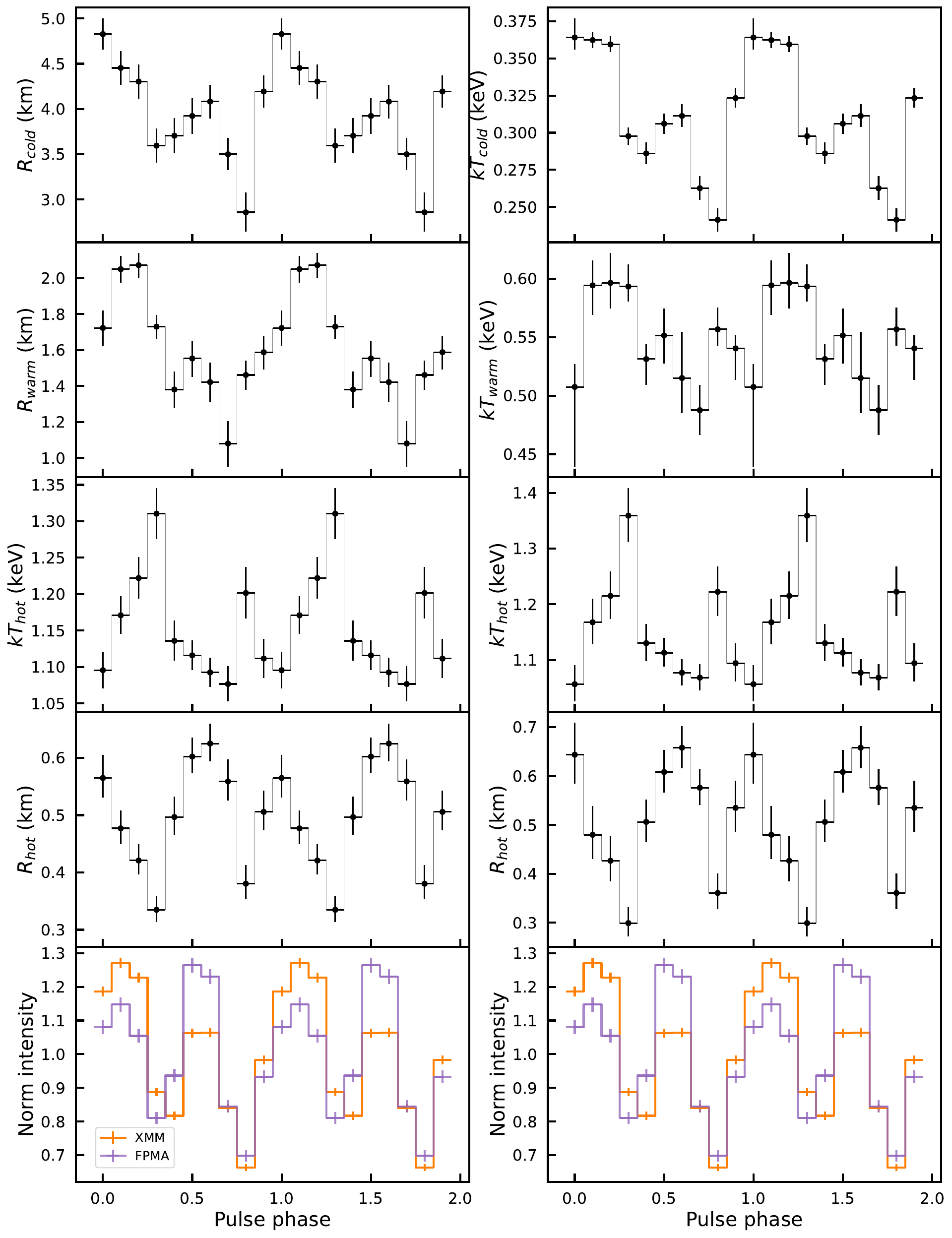}
\caption{Phase evolution of the free spectral parameters, with the \xmm\ (orange) and \nustar/FPMA (purple) pulse profiles (bottom panels). {\it Left}: best-fitting results with $\Gamma$, Norm PL, $kT_{\rm cold}$ and $kT_{\rm warm}$ frozen. {\it Right}: best-fitting results with $\Gamma$, Norm PL, $R_{\rm cold}$ and $R_{\rm warm}$ frozen. }
\label{fig:spec_pps}
\end{figure}

We then carried out a phase-resolved spectral analysis using the simultaneous \xmm/EPIC-pn and \nustar/FPMA datasets (epoch: 2026 July 3). We divided the rotational phase cycle into ten intervals, each with a width of 0.1 in phase, and fitted the phase-averaged model (i.e., an absorbed three-blackbody plus power law model) to the spectra simultaneously. In all the fits, \nh\ was held fixed to the phase-averaged value. We tried different approaches: from allowing all the parameters to vary to allowing only one spectral component to change along the rotational phase. Statistically equivalent good fits were obtained by fixing the power law parameters to the phase-averaged values (see Table\,\ref{tab:broadband_fit}), leaving $kT_{\rm hot}$ and $R_{\rm hot}$ free to vary, and fixing two other parameters, which do not belong to the same spectral component, at the phase-averaged values. 
Figure\,\ref{fig:spec_pps} shows two of these fits. In the left panel, we display the results for the case where $\Gamma$, Norm PL, $kT_{\rm cold}$ and $kT_{\rm warm}$ were frozen ($\chi^2$=1460.2 for 1346 dof). The phase evolution of the radii
closely follows the double-peaked shape of the pulse profile. A similar behaviour for $R_{\rm hot}$ and $kT_{\rm cold}$ is also observed on the right panel, where we report the results for the fit with $\Gamma$, Norm PL, $R_{\rm cold}$ and $R_{\rm warm}$ fixed ($\chi^2$=1477.74 for 1346 dof).

\subsection{Search for short X-ray bursts}
\label{sec:bursts}

We searched the event lists for short bursts using time windows of 5\,ms--2.56\,s starting at each event time, subject to the time resolution of each observing mode. 
We calculated the single-trial Poisson probability of the observed counts, $p_{\rm single}$, and applied a correction for the 7,306,620 valid windows searched, $p_{\rm global}=N_{\rm trial} \times p_{\rm single}$. We retained only events with $p_{\rm global}<10^{-3}$, obtaining a sample of 12 bursts: six detected in the FXT-B TM event lists and six with \nustar{}. For each burst, we subtracted the non-burst baseline measured over the surrounding 100-s interval to estimate their net counts and durations. Table~\ref{tab:bursts} lists their properties, and Fig.~\ref{fig:bursts} shows their light curves. Burst spectra could not be derived due to poor statistics, however the hardness-ratio does not show significant variability (within the large uncertainties) from burst to burst.

The SVOM/ECLAIRs onboard triggering algorithm detected a burst on 2026 August 2 (sb26080210), for which the localisation obtained from the event-by-event data is 1.8 arcmin from \src. During this episode, $\sim100$ photons were collected within a 50\,ms interval. Unfortunately, no simultaneous SVOM observations of \src\ during the bursts detected by EP/FXT and \nustar{} are available. An archival search of SVOM/ECLAIRs observations containing \src\ within its half coded FoV (with a total exposure of $>$27 days up to 2026 July 16) with a time resolution of 100\,ms produced no other burst candidates.

\subsection{Near-infrared upper limits} \label{subsec:ir_counterpart}
Photometric calibration of the GTC/EMIR data was performed using isolated UKIDSS DR11+ stars \citep{Lawrence2007}. For each filter, we fitted $m_{\rm cat}-m_{\rm inst}=ZP+\alpha\,(m_{\rm inst}-m_{\rm piv})+\beta\,[(J-K_{\rm s})-0.8]$, where $m_{\rm cat}$ is the UKIDSS catalogue magnitude, $m_{\rm inst}=-2.5\log_{10}({\rm counts})$, $ZP$ is the photometric zero point, $m_{\rm piv}$ is the median instrumental magnitude of the calibration stars, and the $\alpha$ and $\beta$ terms account for small magnitude- and colour-dependent residuals in the transformation. After outlier rejection, the calibration used 322, 169, and 108 stars in $J$, $H$, and $K_{\rm s}$, yielding rms residuals of 0.04, 0.05, and 0.08 mag, respectively. No point source was detected in the stacked images within the \xmm\ localisation region, down to $3\sigma$ limiting magnitudes of $J>21.5$, $H>21.4$, and $K_{\rm s}>21.0$\,mag (Vega).

\subsection{Radio upper limits} 
\label{subsec:radio_limits}
Data from the Medicina antennas were folded at the X-ray ephemeris using a \texttt{tempo2} \citep{hobbs2006,edwards2006} phase predictor generated for the Medicina site, retaining all 256 channels and 1024 phase bins, and cleaned of radio frequency interference with \textsc{clfd} \citep{morello2019}.
The folded profiles were searched over DM $=20$--500 pc cm$^{-3}$ in steps of 0.5 pc cm$^{-3}$, to include the values predicted by the NE2025 \citep{NE2025} and YMW16 \citep{yao2017} electron density models at $d\simeq3.3$\,kpc (100 and 119 pc cm$^{-3}$). This DM interval includes also the value predicted by the \nh--DM relation of \cite{he13}, which in this case returns 80 pc cm$^{-3}$ assuming a neutral hydrogen column density of $2.4 \times 10^{21}$\,cm$^{-2}$.
We also searched for single dispersed pulses over the same DM range following a standard procedure based on transient detection with \texttt{HEIMDALL} \citep{barsdell2012} and machine learning classification performed by the convolutional neural network of \texttt{FETCH} \citep{agarwal2020}. No periodic signal or single pulse was detected in either epoch.
We adopted a detection threshold of S/N $=10$ for both searches, starting from a $1\sigma$ sensitivity of 1.3\,Jy for a 1\,ms pulse with the two polarisation channels summed, and converted the non-detections into flux density limits with the radiometer equation.
Therefore, for the single-pulse search, assuming a 1\,ms burst width, the fluence limit is $\sim$13\,Jy\,ms, corresponding to an isotropic-equivalent spectral energy of $<1.7\times10^{20}$\,erg\,Hz$^{-1}$ at an assumed distance of 3.3\,kpc.
For the folded search, assuming a 10\% duty cycle, the upper limits on the mean flux density averaged over the rotation are 1.4 and 1.2\,mJy for the two epochs and 0.9\,mJy for the two combined (0.39, 0.62 and 1.4\,mJy for assumed duty cycles of 2\%, 5\% and 20\%).


FAST data, given their exquisite sensitivity, were searched in a conservative DM range of 0$-$1500 \({\rm pc\,cm^{-3}}\). We searched for periodic radio pulsations using a GPU-accelerated pulsar-search pipeline developed on the basis of PRESTO \citep{2011ascl.soft07017R}. Both fast Fourier transform (FFT) and fast-folding algorithm (FFA) searches were performed. No convincing periodic radio signal with \(S/N>7\) was found near the X-ray pulsation period reported in \S\ref{subsec:timing}.
We then estimated upper limits on the mean pulsed flux density $S_{\rm min}$ using the radiometer equation  \citep{2004hpa..book.....L,2025ApJ...979..122B},

\[
S_{\rm min}
=
\frac{\beta\left(S/N_{\min}\right)T_{\rm sys}}
{G\sqrt{n_{\rm p}t_{\rm obs}\Delta f}}
\sqrt{\frac{W}{P-W}},
\]
where \(\beta\) is the digitization-loss factor, \(T_{\rm sys}\) is the system temperature, including receiver and sky temperatures. \(G\) is the telescope gain, \(n_{\rm p}\) is the number of summed polarizations, \(t_{\rm obs}\) is the on-source integration time, \(\Delta f\) is the effective observing bandwidth, \(P\) is the spin period, and \(W\) is the effective pulse width. We adopted \(\beta = 1\), with a signal-to-noise ratio threshold of 7 and an assumed pulse duty cycle of 10\%. Following \cite{2020RAA....20...64J}, both the telescope gain, \(G\), and the receiver temperature, depend on the observing zenith angle, \(\theta_{\rm ZA}\). We therefore evaluated these quantities for each observation and list the corresponding gain, system temperature, and sensitivity limits in Table~\ref{tab:fast_limits}.

We also searched the data for single pulses using TransientX \citep{2024A&A...683A.183M} and using \texttt{single$\_$pulse$\_$search$.$py} from the PRESTO software. No credible astrophysical single-pulse or FRB-like event with \(S/N>9\) was confirmed in any of the two observations. The corresponding upper limit on the peak flux density of a single pulse was calculated as
\[
S_{\rm SP}
=
\frac{\beta\left(S/N_{\min}\right)T_{\rm sys}}
{G}
\sqrt{\frac{1}{n_{\rm p}\Delta f W}}~.
\]

\begin{table}[htbp]
  \centering
  \caption{FAST observation parameters and upper limits.}
  \label{tab:fast_limits}
  \small
  \setlength{\tabcolsep}{3pt}
  \begin{tabular}{@{}lcccc@{}}
    \hline
    \hline
    Date & $T_{\rm sys}$ & $G$ & $S_{\rm min}$ & $S_{\rm sp}$ \\
         & (K) & ($\mathrm{K\,Jy^{-1}}$)
         & ($\mu$Jy) & (mJy) \\
    \hline
    July 4
      & $28.34 \pm 0.11$
      & $15.55 \pm 0.16$
      & $2.30 \pm 0.03$
      & $1.64 \pm 0.02$ \\
    August 2
      & $28.23 \pm 0.12$
      & $15.63 \pm 0.16$
      & $2.26 \pm 0.03$
      & $1.63 \pm 0.02$ \\
    \hline
  \end{tabular}
\end{table}
All other parameters were adopted from the values listed in Table ~\ref{tab:fast_limits}, with the exception of the detection threshold and pulse width, for which we assumed \((S/N)_{\min}=9\) and \(W=100\) ms, respectively.

\subsection{Distance estimate}
\label{sec:distance}
We obtained an indicative distance constraint by sampling the three-dimensional H\,{\sc i} and H$_2$ reconstruction of \citet{Soding2025} along the line of sight to \src. In the face-on Galactic projection, this line of sight intersects the Perseus and Outer spiral arms at heliocentric distances of $2.2$--$3.6$\,kpc and $6.9$--$7.9$\,kpc, respectively. At the source Galactic latitude, these correspond to heights of $z=d\sin b\simeq-0.17$ to $-0.29$\,kpc for the Perseus arm and $-0.55$ to $-0.62$\,kpc for the Outer arm. The hydrogen-density profile reconstructed along the line of sight exhibits a prominent enhancement at $d\simeq3.3$\,kpc, within the Perseus-arm interval, whereas no comparable enhancement is present within the Outer-arm interval. The reconstructed gas distribution thus favours an association with the Perseus arm, although an inter-arm location along the same line of sight cannot be ruled out.

\section{Discussion} \label{sec:discussion}

\src\, displays the main phenomenological properties of an active magnetar. Its spin period of $P\simeq6$~s, relatively large spin-down rate, short X-ray bursts, luminous and decaying X-ray emission, broadband spectrum comprising multiple thermal components and a hard X-ray tail are all characteristic of the magnetar population \citep[e.g.][]{KaspiBeloborodov2017,cotizelati18,esposito21,rea26}.
Using the average spin-down rate measured during the present monitoring campaign, we infer $B_{\rm dip}\simeq1.3\times10^{14}$~G, a characteristic age of $\tau_{\rm c}\simeq34$~kyr, and a rotational energy-loss rate of $\dot E\simeq5.1\times10^{32}$~erg~s$^{-1}$. We caution, however, that the structured timing residuals might indicate additional timing noise and/or torque variability. The measured $\dot P$, and consequently the quantities derived from it, should therefore be regarded as representative of the present observing interval rather than as precise measurements of the long-term secular evolution. Variations of the spin-down torque are commonly observed during magnetar active states and can be associated with changes in the current-carrying magnetosphere \citep[e.g.][]{KaspiBeloborodov2017}.

The radiative evolution of \src\ is also broadly consistent with that observed during other magnetar outbursts \citep[e.g.][]{cotizelati18}. At the adopted distance
of 3.3~kpc, the 0.5--30~keV luminosity decreased from $\simeq1.2\times10^{35}$ to $\simeq5.7\times10^{34}$~erg~s$^{-1}$ during the first month after discovery. The densely sampled EP/FXT light curve can be approximately described by an exponential decay with an $e$-folding timescale of $\tau\simeq31$~d. Decay timescales ranging from weeks to months are commonly measured during magnetar
outbursts, whose light curves frequently display exponential, power-law, or multi-component decays \citep[e.g.][]{cotizelati18}. In the case of \src, the large reduced $\chi^2$ of the single-exponential fit, $\chi^2_\nu\approx7.0$, indicates that this timescale should primarily be regarded as an empirical characterisation of the decline at the beginning of the outburst, rather than as evidence for a strictly exponential physical relaxation. 
A future study of the whole outburst decay, when the source will reach quiescence, will shed more light on the final decay timescale and cooling evolution \citep[see e.g.][for a theoretical study of outburst evolution]{DeGrandis2025}. 

Interestingly, the luminosity decline is accompanied by a decrease in the inferred sizes of the thermally emitting regions, whereas the temperatures do not exhibit an equally clear monotonic evolution. In the broadband fits, the radii associated with the warm and hot thermal components decrease by factors of approximately 1.6 and 1.3, respectively, during the first month. A reduction of the effective emitting area during the decay has been observed in several transient magnetars and is qualitatively expected if the outburst predominantly heats a localised portion of the neutron-star surface \citep[e.g.][]{cotizelati18}. It also arises naturally in the untwisting-magnetosphere scenario, in which electric currents flowing through a bundle of twisted magnetic field lines heat the stellar surface at their footprints. As the magnetospheric twist dissipates, the current-carrying bundle and its associated heated footprint contract \citep{beloborodov09}. Alternatively, energy deposited in the crust during the activation can subsequently diffuse towards the surface and deeper layers, producing a gradually fading thermal outburst \citep[e.g.][]{degrandis22, DeGrandis2025}. These mechanisms are not mutually exclusive, and the present observations do not uniquely distinguish between them. Nevertheless, the evolution observed in \src\, suggests that a reduction of the effective emitting area plays an important role in driving the early luminosity decay.

The pulse properties provide complementary information on the geometry of the emitting regions. Although the 0.5--10~keV pulse profile remains broadly stable throughout the monitoring campaign, its morphology is strongly energy dependent. The relative strengths of the two main pulse components change substantially between the soft and hard X-ray bands, and the hardness ratio varies over the rotational cycle. The rms pulsed fraction remains at $\simeq20$--25\% below $\sim7$~keV and decreases to approximately 10\% at higher energies. At the same time, the relative contribution of the fundamental and second harmonic changes with photon energy. These properties indicate that the observed pulse profile is unlikely to arise from the simple rotational modulation of a single, spectrally homogeneous emitting region.

This picture is supported by phase-resolved spectroscopy. Different choices of free parameters for the thermal components provide statistically comparable descriptions of the phase-resolved spectra, implying that the present data do not uniquely identify which physical quantities are responsible for the spectral modulation. Nevertheless, clear phase-dependent variations of the thermal components are observed.
In one of the representative parametrisations, the inferred radii of the cold and hot components closely follow the double-peaked pulse
profile. The combination of an energy-dependent pulse morphology and phase-dependent thermal emission therefore points towards a
non-uniform surface-temperature distribution, potentially involving multiple heated regions with different temperatures
and positions. Magnetospheric resonant scattering may further modify the angular and spectral distribution of the emerging
radiation \citep[e.g.][]{lyutikov06, rea08, nobili08}.

\begin{figure}
\centering
\includegraphics[width=0.44\textwidth]{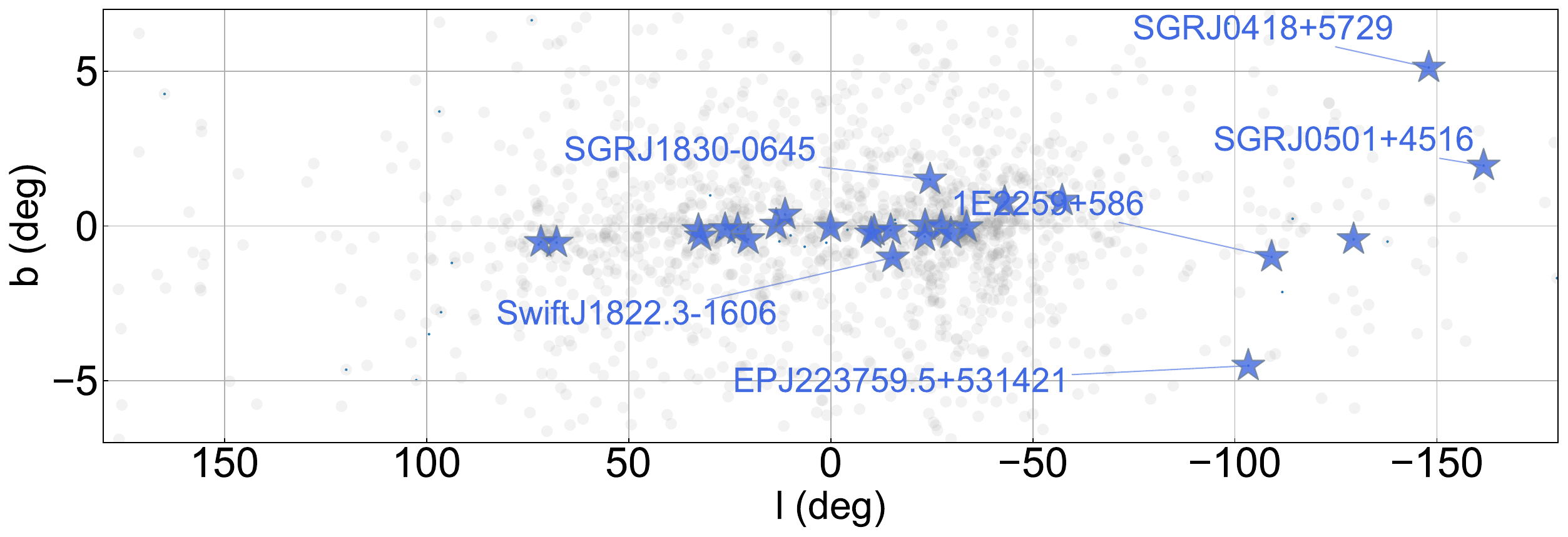}
\includegraphics[width=0.45\textwidth]{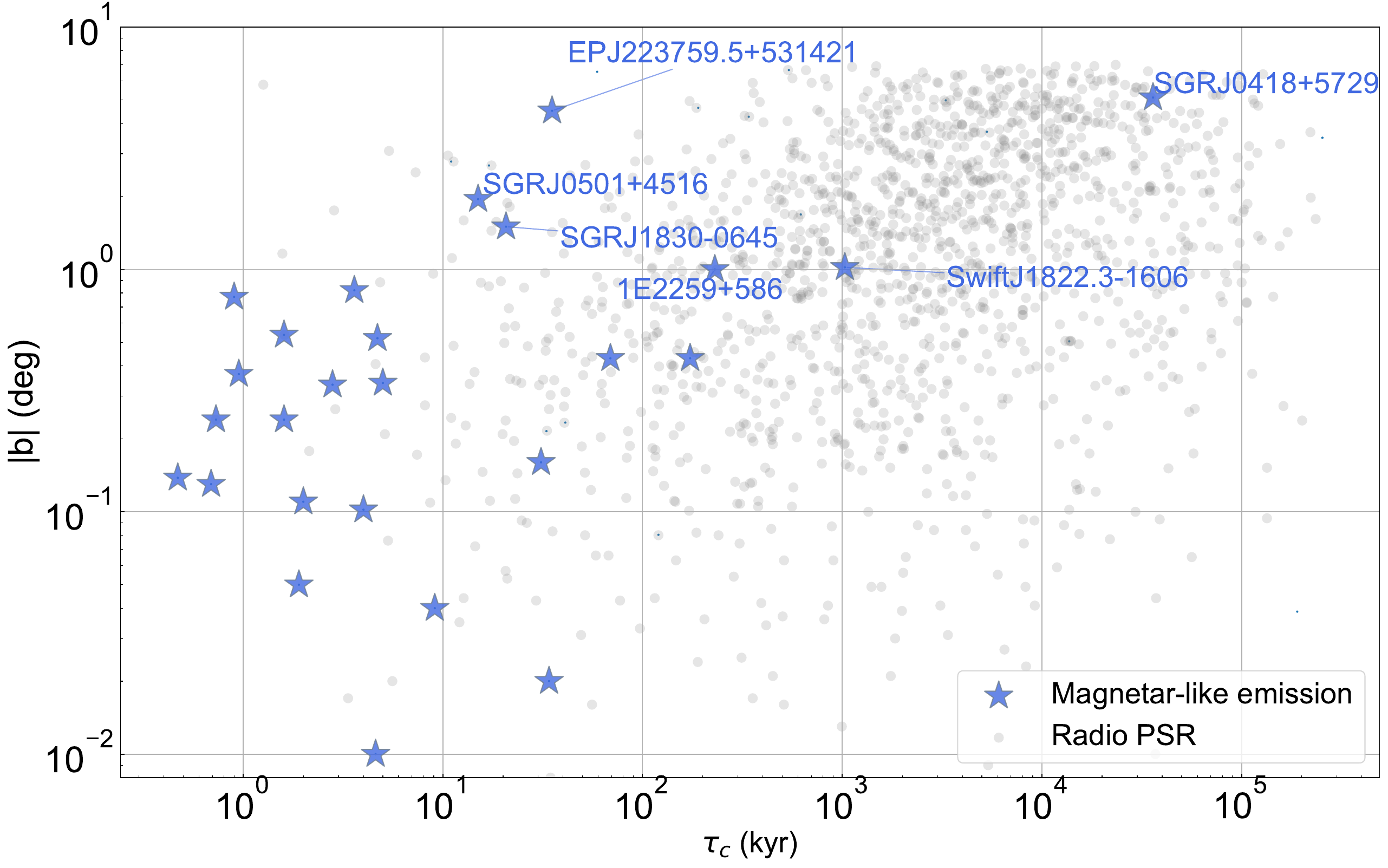}
\includegraphics[width=0.45\textwidth]{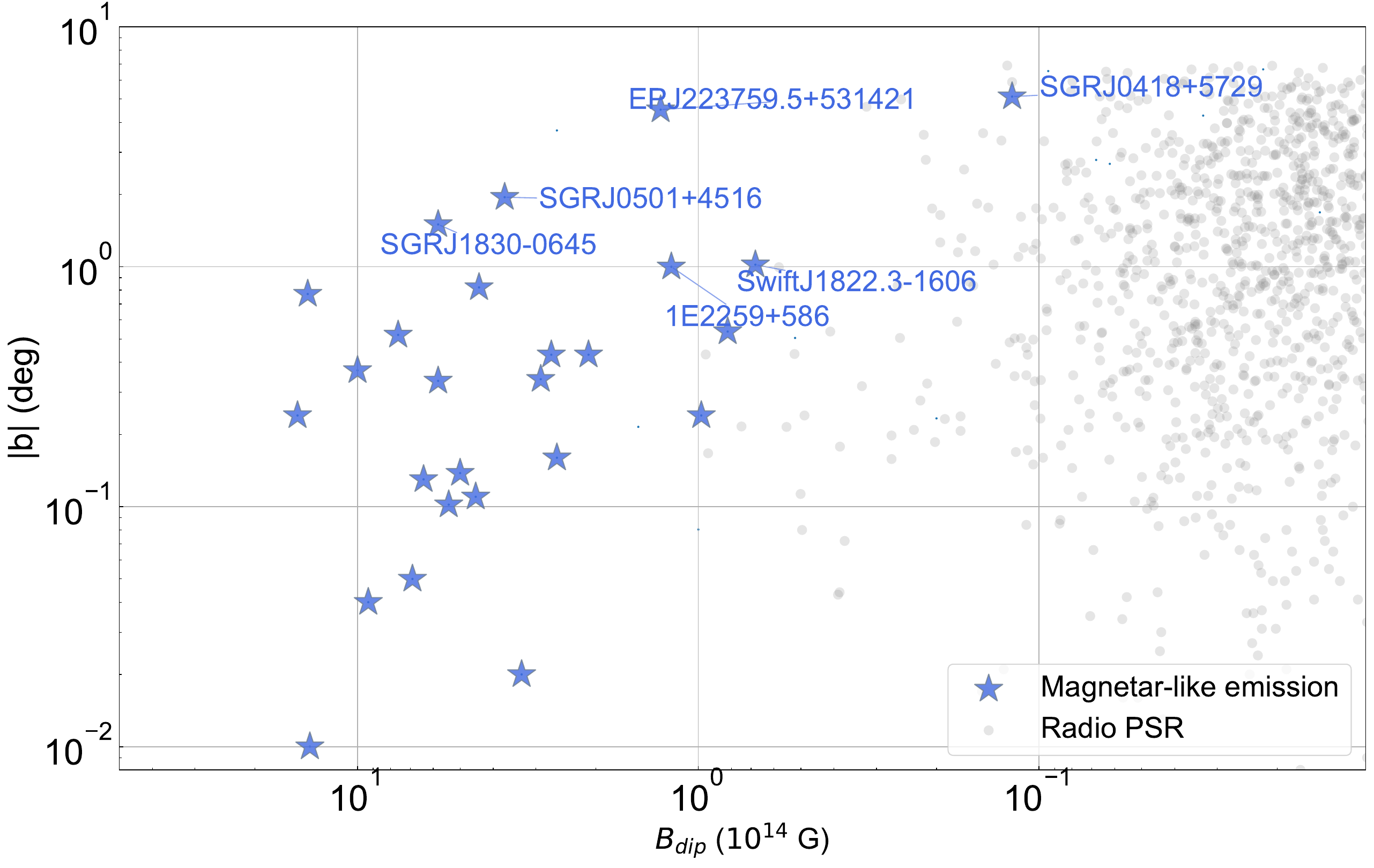}
\caption{Galactic location and spin-down properties of neutron stars displaying magnetar-like emission compared with the radio-pulsar population. Blue stars denote sources with magnetar-like emission, while grey circles represent radio pulsars. Top: Galactic latitude, ($b$), as a function of Galactic longitude, (l). Middle: absolute Galactic latitude, ($|b|$), as a function of characteristic age, ($\tau_{\rm c}$). Bottom: ($|b|$) as a function of the dipolar magnetic-field strength, ($B_{\rm dip}$), inferred under the standard vacuum-dipole assumption. Magnetars with ($|b|>1^\circ$) are labelled. \src\, occupies a comparatively high Galactic latitude for its nominal characteristic age and inferred dipolar magnetic field.}
\label{fig:position}
\end{figure}

It is noteworthy that the broad pulse morphology remains comparatively stable despite the substantial decline in luminosity and the decrease of the inferred emitting areas. This may suggest that the overall geometry of the active regions is preserved during the first months of the outburst, while their characteristic sizes, temperatures and luminosities evolve. Continued monitoring as \src\,
approaches quiescence will be important for determining whether this geometrical stability persists at later stages of the decay.

The detection of emission extending into the hard X-ray band provides a further connection with the broader magnetar population. A spectral model consisting only of two thermal components leaves structured residuals above $\sim10$~keV, whereas the addition of a power-law component provides a satisfactory broadband description at most epochs. The inferred photon index, $\Gamma\sim1.7$--1.9, lies within the range commonly observed for the hard non-thermal components of magnetars \citep{gotz06, enoto17}. Such emission is generally associated with magnetospheric particle populations and the reprocessing of thermal photons in the strongly magnetised magnetosphere. In particular, relativistic particle flows sustained by twisted magnetic field lines provide a possible mechanism for producing hard X-ray radiation \citep{beloborodov13}. The additional cool thermal component required by the high-quality \xmm\ plus \nustar\ spectrum obtained on 2026 July 3 may instead trace emission from a larger fraction of the neutron-star surface.

The short X-ray bursts detected from \src\, independently strengthen its magnetar nature. Our search identifies 12 significant events in the EP/FXT and \nustar\ datasets, with durations ranging from a few tens to $\sim100$~ms, in addition to the bursts detected by other facilities. Their occurrence over several weeks demonstrates that bursting activity continued while the persistent X-ray luminosity was declining, a behaviour commonly observed during magnetar active states \citep[e.g.][]{KaspiBeloborodov2017,esposito21}. Interestingly, \src\ was initially identified through its enhanced persistent soft-X-ray emission, rather than through one of these bursts, which shows the capabilities of \ep\, in catching magnetar outbursts as a soft X-ray monitor.

The archival \swift\ observations indicate that \src\, was substantially fainter before the present activation. No X-ray emission was detected from the source position in the stacked 2017 XRT observations, down to a $3\sigma$ upper limit of 0.014~counts~s$^{-1}$ in the 0.3--10~keV band. Assuming a quiescent emission modelled by a blackbody with $kT$=0.3\,keV, and the measured $N_{\rm H}=2.4\times10^{21}$\,cm$^{-2}$, this limit results in a quiescent unabsorbed flux of $<3\times10^{-13}$\,\flux\ that at a distance of 3.3\,kpc corresponds to a quiescent luminosity $L_{qui}<4\times10^{32}$\,\lum . A deep X-ray observation after the source has returned to quiescence will be particularly valuable in constraining the true quiescent emission of \src. Magneto-thermal evolutionary models predict that the persistent thermal luminosity decreases as the neutron star ages and its magnetic field evolves \citep[e.g.][]{Vigano2013,Ascenzi2024}. A measurement of the quiescent luminosity could consequently provide useful complementary information on the evolutionary state of \src. 

The Galactic location of \src\, is potentially interesting in this context. Along its line of sight, the three-dimensional distribution of Galactic atomic and molecular hydrogen intersects both the Perseus and Outer spiral arms. The reconstructed gas-density profile shows a prominent enhancement at $d\simeq3.3$~kpc, within the Perseus-arm distance range, whereas no similarly strong enhancement is present over the distance interval associated with the Outer arm. We therefore adopt 3.3~kpc as the fiducial distance, while noting that an inter-arm location cannot presently be excluded. At this distance, the Galactic latitude $b=-4.562^\circ$ corresponds to a vertical height of approximately $z\simeq-0.26$~kpc. An association with the more distant Outer arm would instead place the source at $|z|\sim0.6$~kpc.

The relatively large displacement from the Galactic mid-plane is noteworthy for an object belonging to a population generally associated with young massive-star progenitors. It could indicate a non-negligible natal velocity, a relatively evolved system, or a birthplace itself displaced from the Galactic mid-plane (see Fig.\,\ref{fig:position}). For example, if a plausible young stellar association or cluster can be identified in the future, near the projected trajectory of \src, its angular separation and assumed age could be used to estimate the transverse velocity required to reach the present position. This  relatively large $|b|$, the faint pre-outburst state, and the nominal characteristic age are jointly compatible with an evolved magnetar, but none of these quantities independently provides a secure age estimate. 
Assuming an age comparable to its characteristic age (34\,kyr), assuming an extreme 1000\,km/s transverse velocity, the source should have travelled at most $\sim$35pc, that at a 3.3\,kpc distance would mean being born at most at $b\sim-3.9^{\deg}$, still far below the Galactic plane.

No counterpart to \src\, was detected at near-infrared wavelengths. The GTC/EMIR observations yield $3\sigma$ limits of $J>21.5$, $H>21.4$, and $K_{\rm s}>21.0$ mag (Vega) within the refined \xmm\ error circle. Near-infrared counterparts of magnetars span a wide range of luminosities and can vary substantially during and after X-ray outbursts \citep[e.g.][]{esposito21}. The current non-detection is therefore not unexpected, and deeper observations, particularly if obtained at different stages of the X-ray decay, may still reveal a counterpart. Converting the present limits into extinction-corrected luminosity constraints may also allow a more direct comparison with the infrared-to-X-ray luminosity ratios measured in other magnetars.

Similarly, the radio observations reveal neither periodic emission at the X-ray spin period nor individual dispersed radio bursts. Our deepest upper limits are derived by the FAST telescope, resulting in a mean 1.0--1.4-GHz flux density of $\sim2.3$~$\mu$Jy (see Table \ref{tab:fast_limits}). Radio pulsations have been detected from only a minority of known magnetars and can be strongly variable or transient, in most of the cases appearing after episodes of X-ray activity \citep[e.g.][]{camilo06}. The non-detection of radio emission from \src, therefore, does not by itself place a strong constraint on the underlying emission mechanism. In particular, geometrical beaming and temporal variability can prevent detection even if coherent radio emission is produced. Continued radio observations remain worthwhile, especially following renewed bursting activity or significant changes in the X-ray spectral and timing properties, or polarization changes (see also Taverna et al. 2026, submitted).

In conclusion, \src\ is a new magnetar discovered by \ep\ during an outburst and subsequently identified as a magnetar through the detection of typical short X-ray bursts. Its position far below the Galactic plane, at a high Galactic latitude of $|b|\simeq4.56^\circ$, is difficult to reconcile with its characteristic age of 34\,kyr, even assuming an unusually high natal velocity. This may indicate that the source was born well away from the Galactic midplane, that its true age differs substantially from its characteristic age, or that its distance or evolutionary history is more complex than currently inferred. Further optical and infrared observations, particularly proper-motion measurements and searches for a possible birth environment, are needed to constrain the origin of this intriguing source. The detection of the event from the earliest stages of its outburst provides an opportunity to characterise the onset and evolution of a magnetar outburst and demonstrates the potential of \ep's wide-field monitoring to uncover previously quiescent members of the Galactic magnetar population.

\begin{acknowledgements}
 This work is based on the data obtained with the \ep, a space mission led by the Chinese Academy of Sciences, in collaboration with the European Space Agency, the Max Planck Institute for Extraterrestrial Physics (Germany), and the Centre National d'Études Spatiales (France); the \nustar\ mission, a project led by the California Institute of Technology, managed by the Jet Propulsion Laboratory, and funded by NASA; \xmm, an ESA science mission with instruments and contributions directly funded by ESA Member States and the USA (NASA); the Space-based multi-band Variable Objects Monitor (SVOM), a joint Chinese-French mission led by the Chinese National Space Administration (CNSA), the French Space Agency (CNES), and the Chinese Academy of Sciences (CAS).
This work is based on observations made with the Gran Telescopio Canarias (GTC), installed at the Spanish Observatorio del Roque de los Muchachos of the Instituto de Astrofisica de Canarias on the island of La Palma, under programme GTCMULTIPLE3B-26A (PI: F.~Coti Zelati).
This work is based on data obtained with the instrument EMIR, built by a Consortium led by the Instituto de Astrofísica de Canarias. EMIR was funded by GRANTECAN and the National Plan of Astronomy and Astrophysics of the Spanish Government. We thank the GTC staff for their prompt support and execution of these observations. This work made use of the data from FAST (Five-hundred-meter Aperture Spherical radio Telescope). FAST is a Chinese national mega-science facility, operated by National Astronomical Observatories, Chinese Academy of Sciences. We thank FAST for approving our observing proposal. This work is supported by the National Key R\&D Program of China (2021YFA0718504). This work is supported by the National Natural Science Foundation of China (Grant Nos. 12333004 and 12433005), and the Strategic Priority Research Program of the Chinese Academy of Sciences (Grant No. XDB0550200).
AG acknowledges the PhD program in Space Science and Technology at the University of Trento, Cycle XXXIX, with the support of a scholarship financed by the Ministerial Decree no. 118 of 2nd March 2023, based on the NRRP - funded by the European Union - NextGenerationEU -- CUP E66E23000110001.

\end{acknowledgements}

\bibliographystyle{aa}
\bibliography{biblio}

\newpage

\begin{appendix}

\section{Observation log and supplementary results}

\begin{figure*}
\centering
\includegraphics[width=\textwidth]{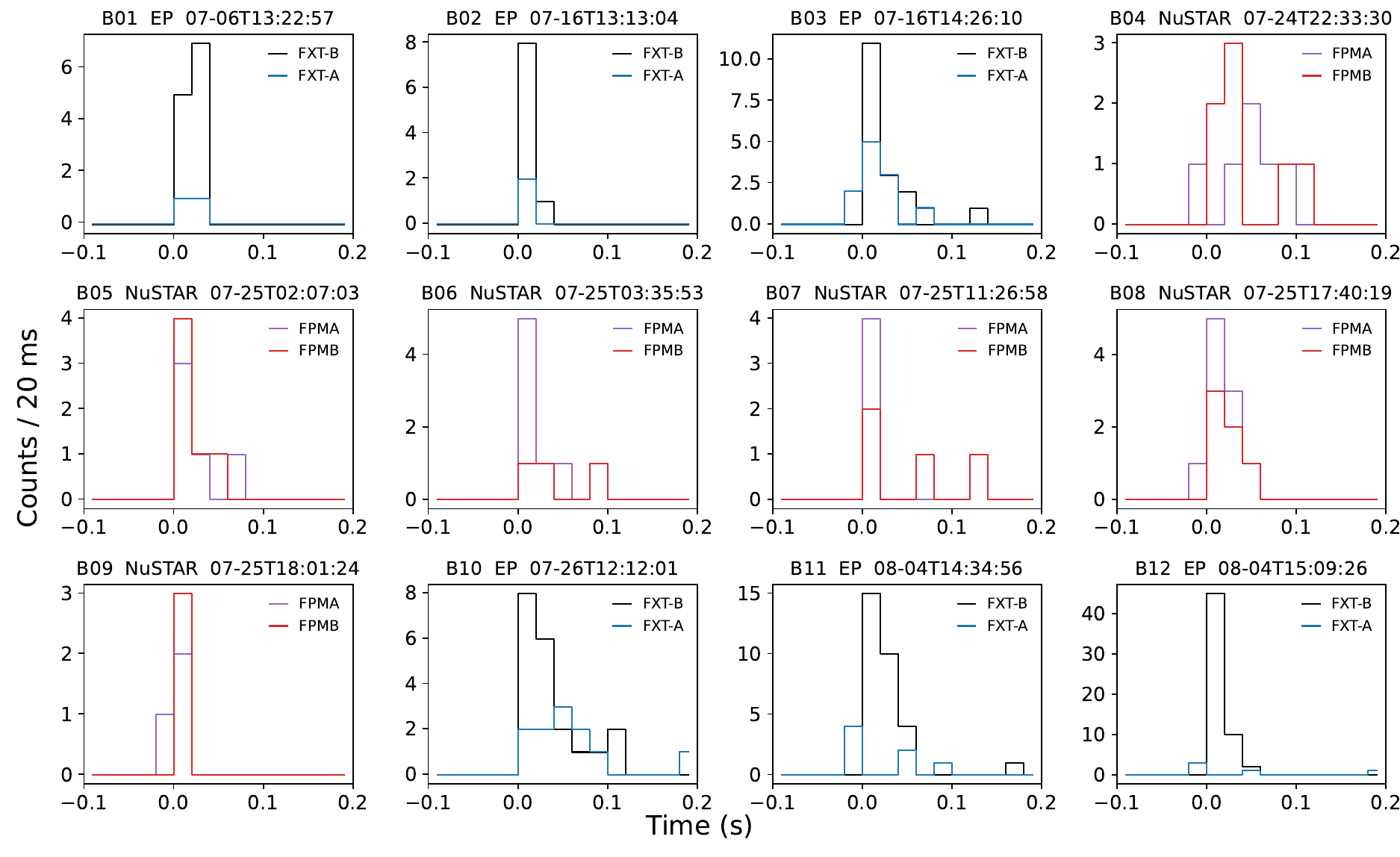}
\vspace{-0.5cm}
\caption{Light curves of the 12 short X-ray bursts detected in our dataset.
Black and blue show EP FXT-B and FXT-A, respectively. Purple and red show
\nustar\ FPMA and FPMB, respectively.}
\label{fig:bursts}
\end{figure*}

Table~\ref{tab:log} provides the complete log of the X-ray observations analysed in this work.
Tables~\ref{tab:broadband_fit} and \ref{tab:fxt_spectral_fit} report the detailed results of the broadband and EP/FXT spectral fits, respectively.
Table~\ref{tab:bursts} lists the measured properties of the detected short bursts, while Figure~\ref{fig:bursts} shows their light curves.

\begin{table*}
\centering
\caption{Journal of the X-ray observations of \src\ presented in this work {\bf Add SVOM/MXT}}
\label{tab:log}
\begin{tabular}{lcccc}
\hline\hline
Telescope & Mode & ObsID & Start -- End Time (UTC) & Exposure \\
&      &       & (YYYY-MM-DD HH:MM:SS)    & (ks) \\
\hline
Swift/XRT & PC & 07006980001 & 2017-02-28 18:47:26 -- 2017-02-28 18:48:40 & 0.06 \\
Swift/XRT & PC & 07006980002 & 2017-03-04 23:18:13 -- 2017-03-04 23:26:52 & 0.49 \\
\hline
EP/FXT & FF+FF & 08500000639 & 2026-06-29 13:02:00 -- 2026-06-29 13:44:48 & 2.6 \\
EP/FXT & FF+FF & 08500000640 & 2026-06-30 14:34:48 -- 2026-06-30 15:17:33 & 2.1 \\
NuSTAR/FPMA+FPMB &  & 81102313002 & 2026-06-30 21:47:39 -- 2026-07-01 08:14:34 & 21.0--20.8 \\
EP/FXT & PW+TM & 08500000641 & 2026-07-02 01:42:21 -- 2026-07-02 04:01:02 & 4.0 \\
NuSTAR/FPMA+FPMB & & 81102313004 & 2026-07-03 03:46:02 -- 2026-07-03 14:14:23 & 22.2--21.9 \\
XMM--Newton/MOS1+MOS2+pn & SW/FF/SW  & 0963464101 & 2026-07-03 22:36:59 -- 2026-07-04 02:22:06 & 12.0--12.8 \\
EP/FXT & PW+PW & 08500000642 & 2026-07-04 20:45:53 -- 2026-07-04 21:20:50 & 2.1 \\
SVOM/MXT & PHOTON & 1140860160 & 2026-07-05 15:25:07 -- 2026-07-05 17:31:15 & 3.5 \\
IXPE & GPD & 05250701 & 2026-07-06 09:29:50 -- 2026-07-10 07:09:15 & 200.2 \\
EP/FXT & PW+TM & 08500000643 & 2026-07-06 11:05:05 -- 2026-07-06 13:24:17 & 5.1 \\
SVOM/MXT & PHOTON & 1140860165 & 2026-07-07 17:32:09 -- 2026-07-07 22:52:34 & 2.6 \\
EP/FXT & PW+TM & 08500000644 & 2026-07-07 22:12:43 -- 2026-07-08 00:32:18 & 5.2 \\
SVOM/MXT & PHOTON & 1140860167 & 2026-07-08 14:32:57 -- 2026-07-08 16:39:40 & 3.6 \\
EP/FXT & PW+TM & 08500000647 & 2026-07-10 14:04:52 -- 2026-07-10 16:33:33 & 6.0 \\
EP/FXT & PW+TM & 08500000648 & 2026-07-12 12:23:15 -- 2026-07-12 14:12:38 & 4.4 \\
EP/FXT & PW+TM & 06800001625 & 2026-07-14 18:50:13 -- 2026-07-14 20:50:03 & 4.7 \\
EP/FXT & PW+TM & 06800001634 & 2026-07-16 12:23:36 -- 2026-07-16 14:52:49 & 6.3 \\
EP/FXT & PW+TM & 06800001638 & 2026-07-18 15:30:10 -- 2026-07-18 16:22:23 & 3.1 \\
NuSTAR/FPMA+FPMB &  & 81102313006 & 2026-07-20 00:46:13 -- 2026-07-20 03:16:20 & 6.2--6.1 \\
EP/FXT & PW+TM & 06800001643 & 2026-07-20 15:24:05 -- 2026-07-20 17:51:23 & 6.1 \\
EP/FXT & PW+TM & 06800001648 & 2026-07-22 20:06:03 -- 2026-07-22 22:31:40 & 5.7 \\
EP/FXT & PW+TM & 06800001652 & 2026-07-24 21:38:38 -- 2026-07-25 00:00:14 & 5.3 \\
NuSTAR/FPMA+FPMB &  & 81102313008 & 2026-07-24 22:19:36 -- 2026-07-25 19:53:56 & 41.1--40.7 \\
EP/FXT & PW+TM & 06800001657 & 2026-07-26 11:52:02 -- 2026-07-26 14:18:25 & 6.0 \\
EP/FXT & PW+TM & 06800001673 & 2026-07-30 03:38:52 -- 2026-07-30 06:05:01 & 5.7 \\
EP/FXT & FF+FF & 06800001684 & 2026-08-02 13:02:31 -- 2026-08-02 13:52:47 & 3.0 \\
EP/FXT & FF+TM & 06800001687 & 2026-08-04 12:55:13 -- 2026-08-04 15:21:13 & 6.0 \\
EP/FXT & FF+TM & 06800001688 & 2026-08-06 19:10:54 -- 2026-08-06 21:36:52 & 5.9 \\
EP/FXT & FF+TM & 06800001700 & 2026-08-11 06:11:17 -- 2026-08-11 08:27:34 & 5.2 \\
EP/FXT & FF+TM & 06800001706 & 2026-08-12 14:07:23 -- 2026-08-12 16:30:24 & 5.7 \\
EP/FXT & FF+TM & 11900855040 & 2026-08-18 05:48:49 -- 2026-08-18 09:36:54 & 6.0 \\
EP/FXT & FF+TM & 11900859776 & 2026-08-22 05:54:24 -- 2026-08-22 10:36:43 & 5.7 \\
\hline
\end{tabular}
\tablefoot{
PC: Photon Counting mode; FF: Full Frame mode; PW: Partial Window mode; TM: Timing mode; SW: Small Window mode; PHOTON: MXT photon-event acquisition mode. \nustar\ and \xmm\ exposure ranges reflect the separate detector event files.
}
\end{table*}


\begin{table*}
\centering
\caption{Results of the joint fit for the broadband spectra of \src.}
\label{tab:broadband_fit}
\scriptsize
\setlength{\tabcolsep}{3pt}
\resizebox{\textwidth}{!}{%
\begin{tabular}{lccccccccc}
\hline\hline
Epoch
& $kT_{\rm cold}$
& $R_{\rm cold}^{a}$
& $kT_{\rm warm}$
& $R_{\rm warm}^{a}$
& $kT_{\rm hot}$
& $R_{\rm hot}^{a}$
& $\Gamma$
& Norm PL
& Flux$^{b}$ (Obs./Unabs.) \\
& (keV)
& (km)
& (keV)
& (km)
& (keV)
& (km)
&
& ($10^{-4}$\,keV$^{-1}$\,cm$^{-2}$\,s$^{-1}$)
& ($10^{-11}$\,\flux) \\
\hline
2026 June 30
& --
& --
& $0.45\pm0.01$
& $2.8\pm0.1$
& $1.09\pm0.01$
& $0.60\pm0.01$
& $1.7\pm0.2$
& $7.6^{+6.2}_{-3.7}$
& $8.2\pm0.1$ / $9.1\pm0.1$ \\

2026 July 3$^{c}$
& $0.31\pm0.03$
& $3.9\pm0.4$
& $0.56_{-0.04}^{+0.06}$
& $1.6\pm0.3$
& $1.13\pm0.02$
& $0.52\pm0.04$
& $1.3\pm0.3$
& $1.9^{+1.8}_{-1.0}$
& $7.8\pm0.1$ / $8.7\pm0.1$ \\

2026 July 20
& --
& --
& $0.50\pm0.02$
& $1.6\pm0.1$
& $1.09\pm0.02$
& $0.46\pm0.02$
& $1.9\pm0.2$
& $7.8\pm3.2$
& $4.6\pm0.1$ / $5.1\pm0.1$ \\

2026 July 24
& --
& --
& $0.47\pm0.01$
& $1.7\pm0.1$
& $1.05\pm0.01$
& $0.47\pm0.01$
& $1.8^{+0.2}_{-0.4}$
& $4.4^{+3.4}_{-3.1}$
& $4.0\pm0.1$ / $4.4\pm0.1$ \\
\hline
\end{tabular}%
}
\tablefoot{The fitted hydrogen column density is
$\nh=(2.4\pm0.2)\times10^{21}$\,cm$^{-2}$.
$^{a}$ Radii were calculated assuming a distance of 3.3\,kpc.
$^{b}$ Observed and unabsorbed fluxes were calculated over the
0.5--30\,keV energy range. $^{c}$ For only this epoch, the soft X-ray coverage is given by \xmm, while for the others by EP. }
\end{table*}

\begin{table*}
\centering
\caption{Results of the spectral fits for the EP/FXT spectra of \src.}
\label{tab:fxt_spectral_fit}
\begin{tabular}{lcccccc}
\hline
\hline
Epoch & $kT_{\rm warm}$ & $R_{\rm warm}^a$ & $kT_{\rm hot}$ & $R_{\rm hot}^a$ & Flux$^b$ (Obs / Unabs) & $\chi^2/{\rm dof}$ \\
      & (keV)            & (km)             & (keV)           & (km)            & ($10^{-11}\,\flux$)    &                       \\
\hline
2026 June 29 & $0.43\pm0.01$ & $3.11^{+0.09}_{-0.08}$ & $1.13\pm0.05$ & $0.53\pm0.05$ & $6.76\pm0.13$ / $7.56\pm0.13$ & $584/439$ \\
2026 June 30 & $0.44^{+0.01}_{-0.02}$ & $2.92^{+0.11}_{-0.10}$ & $1.11^{+0.07}_{-0.06}$ & $0.55\pm0.07$ & $6.60^{+0.16}_{-0.15}$ / $7.36^{+0.16}_{-0.15}$ & $389/384$ \\
2026 July 2 & $0.43^{+0.01}_{-0.02}$ & $3.36^{+0.13}_{-0.12}$ & $1.07^{+0.06}_{-0.05}$ & $0.64^{+0.08}_{-0.07}$ & $7.78\pm0.16$ / $8.72\pm0.16$ & $268/276$ \\
2026 July 4 & $0.48\pm0.01$ & $2.60^{+0.08}_{-0.07}$ & $1.20^{+0.09}_{-0.07}$ & $0.45^{+0.07}_{-0.06}$ & $6.88\pm0.16$ / $7.62\pm0.16$ & $447/392$ \\
2026 July 6 & $0.44\pm0.02$ & $2.76\pm0.10$ & $1.07^{+0.06}_{-0.05}$ & $0.59\pm0.07$ & $6.54\pm0.13$ / $7.27\pm0.13$ & $260/283$ \\
2026 July 7 & $0.43\pm0.01$ & $2.75^{+0.11}_{-0.10}$ & $1.09\pm0.05$ & $0.57^{+0.06}_{-0.05}$ & $6.26\pm0.12$ / $6.94\pm0.12$ & $276/279$ \\
2026 July 10 & $0.40\pm0.02$ & $2.65^{+0.15}_{-0.13}$ & $0.99^{+0.05}_{-0.04}$ & $0.68\pm0.07$ & $5.55\pm0.11$ / $6.15\pm0.11$ & $318/281$ \\
2026 July 12 & $0.41\pm0.02$ & $2.46^{+0.15}_{-0.13}$ & $1.03^{+0.06}_{-0.05}$ & $0.62\pm0.07$ & $5.30\pm0.13$ / $5.85\pm0.13$ & $230/240$ \\
2026 July 14 & $0.40\pm0.02$ & $2.45^{+0.17}_{-0.15}$ & $1.02^{+0.06}_{-0.05}$ & $0.61\pm0.07$ & $4.81\pm0.12$ / $5.31\pm0.12$ & $217/238$ \\
2026 July 16 & $0.39\pm0.02$ & $2.32^{+0.15}_{-0.13}$ & $1.01\pm0.04$ & $0.63\pm0.06$ & $4.76\pm0.10$ / $5.23\pm0.10$ & $274/267$ \\
2026 July 18 & $0.46^{+0.03}_{-0.04}$ & $1.97^{+0.15}_{-0.12}$ & $1.12^{+0.14}_{-0.10}$ & $0.45^{+0.11}_{-0.10}$ & $4.27^{+0.18}_{-0.17}$ / $4.70^{+0.18}_{-0.17}$ & $144/153$ \\
2026 July 20 & $0.44\pm0.03$ & $1.91^{+0.13}_{-0.11}$ & $1.03^{+0.07}_{-0.05}$ & $0.54^{+0.08}_{-0.07}$ & $4.03\pm0.09$ / $4.43\pm0.09$ & $288/251$ \\
2026 July 22 & $0.33\pm0.02$ & $2.57^{+0.25}_{-0.21}$ & $0.95^{+0.04}_{-0.03}$ & $0.66\pm0.05$ & $3.83\pm0.09$ / $4.23\pm0.09$ & $219/233$ \\
2026 July 24 & $0.39\pm0.03$ & $2.01^{+0.18}_{-0.15}$ & $0.96^{+0.06}_{-0.05}$ & $0.60\pm0.08$ & $3.54\pm0.10$ / $3.90\pm0.10$ & $177/218$ \\
2026 July 26 & $0.41\pm0.03$ & $2.00^{+0.16}_{-0.13}$ & $1.01^{+0.07}_{-0.06}$ & $0.52^{+0.08}_{-0.07}$ & $3.34\pm0.09$ / $3.69\pm0.09$ & $235/225$ \\
2026 July 30 & $0.52\pm0.03$ & $1.47^{+0.07}_{-0.06}$ & $1.33^{+0.17}_{-0.13}$ & $0.27\pm0.06$ & $3.37^{+0.13}_{-0.12}$ / $3.66^{+0.13}_{-0.12}$ & $182/204$ \\
2026 August 2 & $0.40\pm0.03$ & $1.74^{+0.12}_{-0.10}$ & $0.98^{+0.06}_{-0.05}$ & $0.50\pm0.06$ & $2.67\pm0.08$ / $2.94\pm0.08$ & $255/223$ \\
2026 August 4 & $0.38\pm0.03$ & $1.88^{+0.17}_{-0.14}$ & $0.99^{+0.05}_{-0.04}$ & $0.50\pm0.05$ & $2.70\pm0.07$ / $2.97\pm0.07$ & $184/192$ \\
2026 August 6 & $0.32\pm0.04$ & $2.18^{+0.35}_{-0.26}$ & $0.83^{+0.05}_{-0.04}$ & $0.66\pm0.08$ & $2.22\pm0.06$ / $2.50\pm0.06$ & $178/180$ \\
2026 August 11 & $0.42\pm0.03$ & $1.62^{+0.15}_{-0.12}$ & $1.12^{+0.09}_{-0.08}$ & $0.35^{+0.06}_{-0.05}$ & $2.32\pm0.08$ / $2.55\pm0.08$ & $176/140$ \\
2026 August 12 & $0.39\pm0.03$ & $1.78^{+0.14}_{-0.12}$ & $1.03^{+0.08}_{-0.07}$ & $0.39\pm0.06$ & $2.11\pm0.07$ / $2.34\pm0.07$ & $157/149$ \\
2026 August 18 & $0.42\pm0.03$ & $1.53^{+0.12}_{-0.10}$ & $1.05^{+0.10}_{-0.08}$ & $0.31\pm0.06$ & $1.64\pm0.06$ / $1.83\pm0.06$ & $147/131$ \\
2026 August 22 & $0.24^{+0.04}_{-0.03}$ & $2.87^{+0.77}_{-0.57}$ & $0.75^{+0.04}_{-0.03}$ & $0.66\pm0.06$ & $1.33\pm0.04$ / $1.52\pm0.04$ & $124/114$ \\
\hline
\end{tabular}
\tablefoot{The spectra were fitted with ${\tt tbabs}\times({\tt bbodyrad}+{\tt bbodyrad})$. The hydrogen column density was fixed at $N_{\rm H}=2.4\times10^{21}\,{\rm cm}^{-2}$. $^a$ Radii are calculated considering a distance of 3.3\,kpc. $^b$ Fluxes are estimated in the 0.5--10\,keV energy range.}
\end{table*}

\begin{table*}
\centering
\caption{Properties of the short X-ray bursts detected from \src}
\label{tab:bursts}
\begin{tabular}{lccccc}
\hline\hline
Burst & Instrument & UTC window start & Net counts & $T_{90}$ & $p_{\rm global}$ \\
&            & (YYYY-MM-DD HH:MM:SS) &            & (ms) & \\
\hline
B01 & EP/FXT & 2026-07-06 13:22:57.960 & $12+2$ & 30  & $1.1\times10^{-9}$ \\
B02 & EP/FXT & 2026-07-16 13:13:04.195 & $8+2$  & 14  & $4.1\times10^{-6}$ \\
B03 & EP/FXT & 2026-07-16 14:26:10.022 & $14+8$ & 22  & $3.2\times10^{-17}$ \\
B04 & \nustar{} & 2026-07-24 22:33:30.672 & $8+8$  & 92  & $7.9\times10^{-12}$ \\
B05 & \nustar{} & 2026-07-25 02:07:03.698 & $6+7$  & 57  & $7.8\times10^{-11}$ \\
B06 & \nustar{} & 2026-07-25 03:35:53.503 & $11+4$ & 85  & $3.2\times10^{-8}$ \\
B07 & \nustar{} & 2026-07-25 11:26:58.827 & $7+5$  & 130 & $3.5\times10^{-5}$ \\
B08 & \nustar{} & 2026-07-25 17:40:19.734 & $31+9$ & 45  & $5.7\times10^{-25}$ \\
B09 & \nustar{} & 2026-07-25 18:01:24.716 & $6+5$  & 18  & $7.9\times10^{-5}$ \\
B10 & EP/FXT & 2026-07-26 12:12:01.081 & $20+10$ & 102 & $1.4\times10^{-21}$ \\
B11 & EP/FXT & 2026-08-04 14:34:56.169 & $29+8$ & 46  & $1.5\times10^{-49}$ \\
B12 & EP/FXT & 2026-08-04 15:09:26.300 & $55+4$ & 29  & $1.6\times10^{-137}$ \\
\hline
\end{tabular}

\tablefoot{
Net counts are reported as FXT-B+FXT-A for EP and FPMA+FPMB for \nustar{}. The quoted $T_{90}$ values are the 5--95 percentile spans.
}
\end{table*}

\section{Galactic line of sight and source-distance constraint}
\label{app:galactic_los}

\begin{figure*}[t]
\centering
\includegraphics[height=0.335\textwidth]{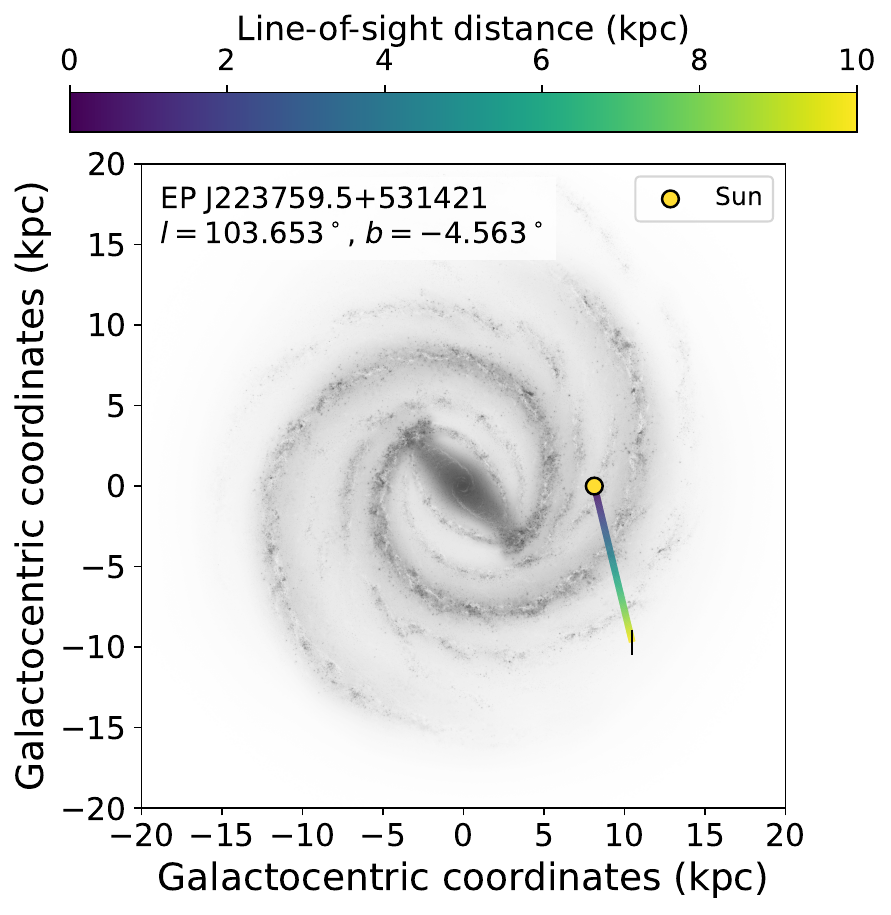}\hfill
\includegraphics[height=0.335\textwidth]{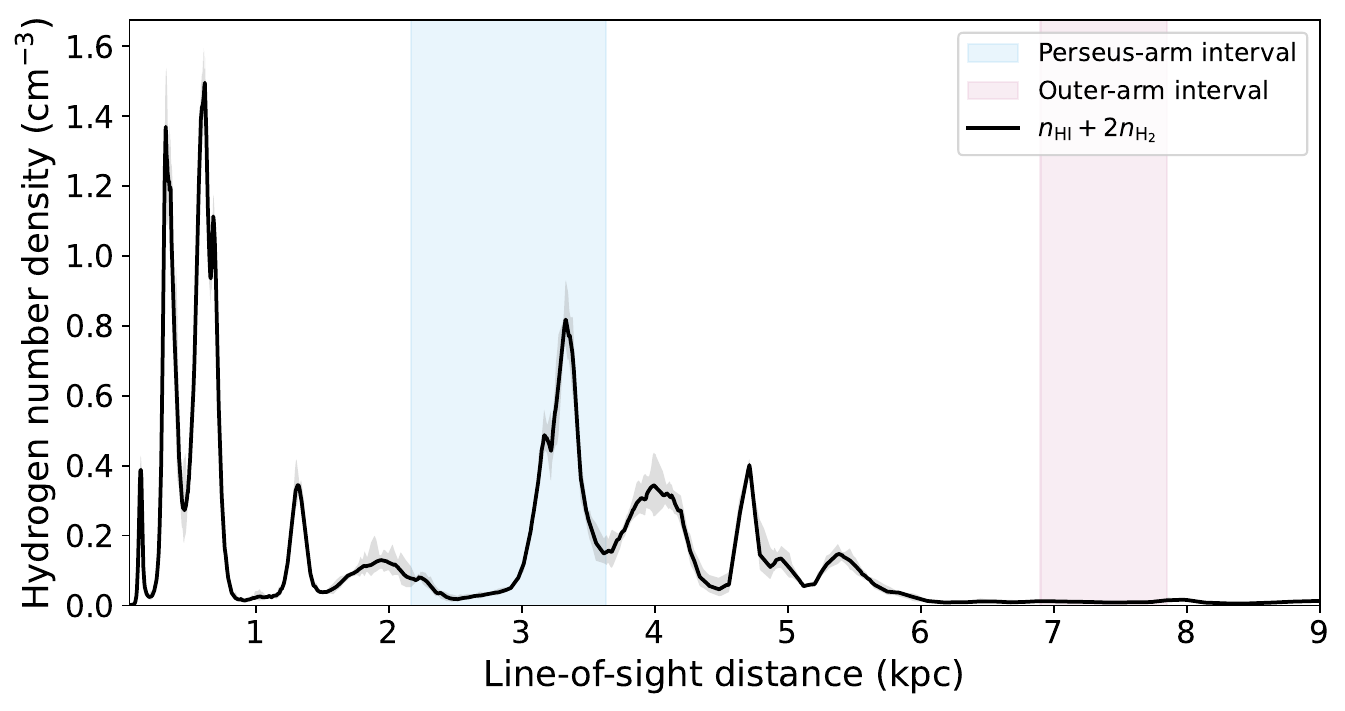}
\caption{Galactic environment along the line of sight to \srcfirst. \textit{Left:} Face-on projection of the Galactic disc showing the line of sight towards \src\ (illustration credit: NASA/JPL-Caltech/R.~Hurt (SSC/Caltech)). The coloured line traces the source sightline outward from the Sun, with colour indicating heliocentric distance. \textit{Right:} Reconstructed total hydrogen number density $n_{\mathrm{H\,I}}+2n_{\mathrm{H}_2}$ as a function of heliocentric distance along the same sightline (see \citealt{Soding2025}). The blue and pink shaded regions indicate the distance intervals over which the sightline crosses the Perseus and Outer spiral arms, respectively. A prominent density enhancement at $d\simeq3.3$\,kpc falls within the Perseus-arm interval, whereas no comparable enhancement is present within the Outer-arm interval.}
\label{fig:galactic_los_density}
\end{figure*}

Figure~\ref{fig:galactic_los_density} provides additional context for the distance estimate discussed in Section~\ref{sec:distance}. The face-on reconstruction in the left panel shows the sightline from the Sun through the Galactic gas distribution, while the one-dimensional profile in the right panel shows how the total hydrogen number density varies with distance along this sightline. Within the distance range associated with the Perseus arm, the strongest density enhancement occurs at $d\simeq3.3$\,kpc. In contrast, the gas density remains comparatively low throughout the distance range associated with the Outer arm.

\end{appendix}

\end{document}